\documentclass[twocolumn]{aastex631}
\DeclareUnicodeCharacter{2212}{-}
\usepackage{amsmath}
\usepackage{enumitem}
\usepackage{graphicx}
\usepackage{multirow}
\def\apjl{Astrophysical Journal Letters}    % Astrophysical Journal, Letters

\def\apjs{ApJS}                                  % Astrophysical Journal, Supplement
\def\aap{Astronomy \& Astrophysics}

\def\gt#1{Georgia Institute of Technology#1 (GeorgiaTech#1)\gdef\gt{GeorgiaTech}}
\def\si#1{Senior Investigator#1 (SI#1)\gdef\si{SI}}
\def\emri#1{Extreme Mass-Ratio Inspiral#1 (EMRI#1)\gdef\emri{EMRI}}
\def\imbh#1{Intermediate Mass Black Hole#1 (IMBH#1)\gdef\imbh{IMBH}}
\def\smbh#1{supermassive black hole#1 (SMBH#1)\gdef\smbh{SMBH}}
\def\bbh#1{binary black hole#1 (BBH#1)\gdef\bbh{BBH}}
\def\pbh#1{primordial black hole#1 (PBH#1)\gdef\pbh{PBH}}
\def\imbhb#1{intermediate mass black hole binary#1 (IMBHB#1)\gdef\imbhb{IMBHB}}
\def\hmns#1{hypermassive neutron star#1 (HMNS#1)\gdef\hmns{HMNS}}
\def\bh#1{black hole#1\gdef\bh{black hole}}
\def\ns#1{neutron star#1 (NS#1)\gdef\ns{NS}}
\def\hmns#1{hyper-massive neutron star#1 (HMNS#1)\gdef\hmns{HMNS}}
\def\nsbh#1{neutron star-black hole#1 (NSBH#1)\gdef\nsbh{NSBH}}
\def\bns#1{binary neutron star#1 (BNS#1)\gdef\bns{BNS}}
\def\gw#1{gravitational wave#1 (GW#1)\gdef\gw{GW}}
\def\eos#1{equation of state#1 (EOS#1)\gdef\eos{EOS}}
\def\gpu#1{graphics processing unit#1 (GPU#1)\gdef\gpu{GPU}}
\def\gr#1{General Relativity#1 (GR#1)\gdef\gr{GR}}
\def\cbc#1{Compact Binary Coalescence#1 (CBC#1)\gdef\cbc{CBC}}
\def\eob#1{Effective-One-Body#1 (EOB#1)\gdef\eob{EOB}}
\def\pnw#1{post-Newtonian#1 (PN#1)\gdef\pnw{PN}}
\def\pmw#1{post-Minkowskian#1 (PM#1)\gdef\pmw{PM}}
\def\hom#1{Higher Order Mode#1 (HOM#1)\gdef\hom{HOM}}
\def\agn#1{Active Galactic Nucleus#1 (AGN#1)\gdef\agn{AGN}}

\def\ligo#1{Laser Interferometer Gravitational-Wave Observatory#1 (LIGO#1)\gdef\ligo{LIGO}}
\def\lisa#1{Laser Interferometer Space Antennae#1 (LISA#1)\gdef\lisa{LISA}}

\newcommand{\rebound}{{\tt REBOUNDx}}

\newcommand{\msol}{{\mathrm{M}_{\odot}}}

\newcommand{\bremss}{{bremsstrahlung}}

\usepackage{xcolor}
\usepackage{comment}
\usepackage[normalem]{ulem}
\usepackage{soul}   % provides \hl{} for highlighting

\newcommand{\GAtech}{\affiliation{School of Physics, Georgia Institute of Technology, Atlanta, Georgia 30332, USA}}
\newcommand{\Phenikaa}{\affiliation{Phenikaa Institute for Advanced Study, Phenikaa University, Hanoi 10000, Vietnam}}
\newcommand{\IU}{\affiliation{Department of Astronomy, Indiana University, Bloomington, IN 47405, USA}}
\newcommand{\IIT}{\affiliation{Physics Department, Technion: Israel Institute of Technology, Haifa 32000, Israel}}

\begin{document}

\title{Scattering of stellar-mass black holes and gravitational wave bremsstrahlung radiation in AGN disks}

\setlist[enumerate]{itemsep=0mm}

\author{Peter Lott}\GAtech\Phenikaa
\author{Christian Faulhaber}
\affiliation{Johns Hopkins University, Baltimore, Maryland 21218, USA}
\author{Joshua Brandt}\GAtech
\author{Gongjie Li}\GAtech
\author{Hareesh Bhaskar}\GAtech\IU\IIT
\author{Laura Cadonati}\GAtech

\begin{abstract}
Dynamics of stellar mass black holes (sBHs) embedded in active galactic nuclei (AGNs) could produce highly eccentric orbits near the central supermassive black hole, leading 
to repeated close encounters that emit gravitational waves 
in the LIGO frequency band. 
Many works have focused on the mergers of sBH in the disk that produce gravitational waves; however, sBHs in hyperbolic orbits also emit gravitational-wave \bremss{} that can be detected by ground-based interferometers like LIGO. 
In this work, we analyze the scattering of sBHs in an AGN disk as they  
migrate inside the disk, focusing on gravitational-wave \bremss{} emission. We determine how the gravitational-wave emission depends on the different parameters of the scattering experiments, such as the mass of the supermassive black hole and the sBH migration rate and mass ratio.
We find that scattering with detectable gravitational-wave \bremss{} is more frequent around lower mass SMBHs ($\sim 10^{5-6}$M$_\odot$). 
We then conduct a suite of Monte Carlo simulations and estimated the rate for ground-based gravitational-wave detections to be in the range of 0.08 - 1194 $\text{Gpc}^{-3} \text{ yr}^{-1}$, depending on migration forces and detection thresholds, with large uncertainties accounting for  variations in possible AGN environments. The expected rate for our {\tt Fiducial} parameters is 3.2 $\text{Gpc}^{-3} \text{ yr}^{-1}$. Finally, we  provide first-principle gravitational wave templates produced by the encounters. 
\end{abstract}

%%%%%%%%%%%%%%%%%%%%%%%%%%%%%%%%%%%%%%
\section{Introduction} \label{sec:intro}
%%%%%%%%%%%%%%%%%%%%%%%%%%%%%%%%%%%%%%

The discovery of gravitational waves has opened new pathways in the exploration of black holes: 
the majority of signals detected by the \ligo{} and Virgo ground-based interferometers originate from merging binary black hole systems~\cite[]{LIGOScientific:2021djp}, and 
the International Pulsar Timing Array presented results consistent with a gravitational wave background originating from inspiraling supermassive black holes (SMBHs)~\cite[]{NANOGrav:2023gor}.
However, binary coalescences are not the only \gw{} events detectable by  current and next-generation detectors~\cite[]{10.1093/mnras/stab2721}. 
In this paper, we explore the scenario of \bbh{} hyperbolic encounters involving fly-by orbits with gravitational wave \textit{bremsstrahlung} emission, 
which could be detected by \ligo{}, Virgo, and KAGRA if the black holes are sufficiently massive and experience close periastron passages~\cite[]{OLeary:2008myb,Capozziello:2008mn,Kocsis_2006,10.1093/mnras/stab2721}.

The potential detection of such events is interesting 
for several reasons. For one, \bbh{} hyperbolic encounters could provide insight into the inner dynamics of stellar clusters or galaxies by constraining black hole populations. 
These events can also probe General Relativity (GR)~\cite[]{PhysRev.136.B1224, 1977ApJ...216..610T}, complementing the current set of tests of GR with gravitational waves~\cite[]{LIGOScientific:2020tif, LIGOScientific:2021sio}.  
Finally, the GW emission profile is dominated by the non-oscillatory linear memory effect in quadrupolar order, with a distinct transition between early and late waveform phases which for equal mass systems may be louder than the  Christodoulou memory effect that is typical of  mergers~\cite[]{Zeldovich:1974gvh, PhysRevD.45.520, Favata:2010zu}. 

Many channels have been proposed to produce \bbh{} interactions as sources of gravitational waves, such as compact objects in isolated binaries \citep[e.g.,][]{dominik_double_2012,kinugawa_possible_2014,belczynski_first_2016,belczynski_origin_2018,giacobbo_merging_2018,spera_merging_2019,bavera_origin_2020}, isolated triple systems \citep[e.g.,][]{Li14_flip, hoang_black_2018,silsbee_lidov-kozai_2017,antonini_binary_2017}, and star clusters \citep[e.g.,][]{oleary_binary_2006,samsing_formation_2014,banerjee_stellar-mass_2017,rodriguez_binary_2016,askar_mocca-survey_2017, di_carlo_merging_2019}. In this paper, we focus on the channel involving \agn{} disks, which have gained considerable attention as a mechanism 
to explain the existence of massive merging black holes \citep[e.g.,][]{Vokrouhlicky98, Cuadra09, McKernan12, Hoang18, Secunda2019, Yang19, Tagawa20, Gerosa21, Bhaskar22}. 
%Nuclear stellar clusters (NSCs), as the densest environments of stars and black holes, coexist with most of the supermassive black holes (SMBHs) \citep{Kormendy13}.
Nuclear stellar clusters (NSCs) are the densest environments of stars and black holes, coexisting with most \smbh{s} \citep[]{Kormendy13}.
Once an AGN disk is formed, some of the stars and black holes in NSCs may be trapped within the disk through star/BH-disk interactions via disk viscous drag and migrate inside the disk \citep{Artymowics93, Kennedy16, Generozov23, Nasim23}. As shown by \citet{Goldreich79}, embedded objects inside disks lead to density perturbations, which in turn exert torque on the embedded objects. Applying this to AGN disks, it has been shown that there are regions where the exerted torque change sign and leading to \textit{migration traps}. Migration traps in the AGN disk can capture multiple isolated black holes, and produce a dense region for binary to be formed through three-body/multi-body scatterings or GW captures \citep{Bellovary16, Leigh18, Peng21, Samsing22}. Even without migration traps, gas drag can lead to black holes binaries and  binary hardening~\citep{Li23, Rowan23, Rozner23, Wang23, Li23}. %Different from the 
Unlike other formation channels, mergers in  AGN disks may produce electromagnetic counterparts due to the gas-rich environment \citep{Stone17, McKernan19, Graham20, Palmese21}. It has been found that black holes mergers in AGN disks occur at a rate of $0.02 - 60 \text{ Gpc}^{−3} \text{yr}^{−1}$ \citep{Tagawa20}.

So far, most studies on the black holes in the AGN disks have focused on the formation and mergers of black holes binaries. However, a large fraction of interactions of the black holes in the disk leads to close encounters and GW \bremss{} emission. For instance, \citet{Li2022} showed using N-body simulations that the distance between two black holes migrating in the AGN disk can become arbitrarily small during a close encounter even without gas dynamical effects, which leads to strong GW \bremss{} emission. A detailed analysis of the dynamical interactions of the close encounters and the resulting GW \bremss{} emission is missing in the literature. Accordingly, in this paper we focus on the scattering of the black holes and the implication on the observation of the resulting GW emission.

Specifically, we study \gw{} \bremss{} emission from \bbh{} hyperbolic encounters as the individual black holes migrate within the AGN disks. 
We include the effects of gas drag in the AGN disk, as well as first and $2.5-$order post Newtonian corrections, and we simulate the dynamical interactions of the black holes using the \rebound{} package \citep{Rein12}. We then construct templates of the GW emission, test whether this emission could be detectable by current ground-based detectors or future space-based observatories, and provide rate estimates from population synthesis models.

The remainder of this paper is structured as follows: in section 2, we describe the physical scenario and methods for our simulations. In section 3, we explore the parameter space of the scattering experiments and characterize how our results depend on the parameters. 
We conduct an MC population synthesis and calculate the occurrence rate of GW \bremss{} radiation in section 4, and we analyze the GW template and implication for GW searches in section 5. We summarize and discuss our results in section 6. 

\section{Methods} \label{sec:style} 

The \gw{} emission from \bbh{} hyperbolic encounters was first studied by~\cite{PhysRev.136.B1224} by analogy with electromagnetic \bremss. Further elucidation came from \cite{Hansen1972},  \cite{1977ApJ...216..610T}, and the post-linear formulation of~\cite{1978ApJ...224...62K}. 
\cite{Capozziello:2008mn} computed \gw{} emission in the quadrupole approximation for four binary cases: circular, elliptical, parabolic and hyperbolic. This has since been {(1)} expanded to {include} post-Newtonian (PN) order by~\cite{Cho:2018upo} {and} %, including 
spin effects~\cite{Cho:2022syn}, {(2)} studied numerically in~\cite{Damour:2014afa, PhysRevD.96.084009, Bae:2020hla, Hopper:2022rwo, Damour:2022ybd, Bae:2023sww}, and {(3)} 
incorporated into analytic waveform models~\cite{Nagar:2018zoe, Ramos-Buades:2021adz, Dandapat:2023zzn}. 

In this study, we investigate the case of BBH hyperbolic encounters embedded in an AGN disk around a SMBH. We included effects of gaseous viscous drag, and we adopted both the 1-PN and 2.5-PN effects for GR effects. We detail the system configuration and specific methods in this section.

\subsection{System Configuration}

\begin{figure} [t]
     \begin{center}
     \includegraphics[width=\columnwidth]{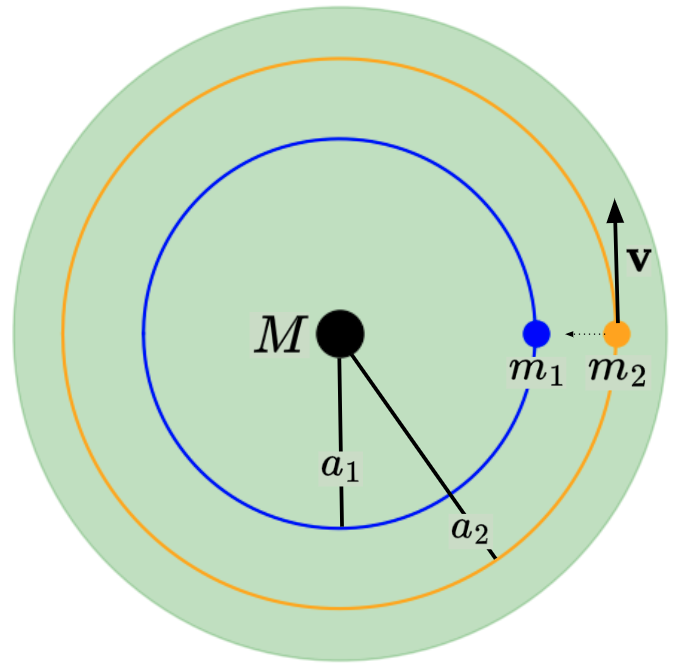}
     \caption{A representation of the system simulated to generate close encounters. The center black dot is the SMBH of mass $M$. The blue and orange orbits represent the inner and perturber stellar mass black holes respectively. The inner black hole has mass $m_1$ and initial semimajor axis $a_1$ (determined by $10^3$ gravitational radii of the SMBH $R_{g, SMBH}$, Eq.~\ref{eq:grav_radius}) and the perturber black hole has mass $m_2$ and semimajor axis $a_2$ (determined such that $a_2 - a_1 = 5$ mutual hill radii $R_{H12}$, Eq.~\ref{eq:mutual-hill-radius}). The perturber black holes has instantaneous velocity $\textbf{v} $ which is needed to calculate drag forces that cause it to migrate inward, represented by the dashed arrow.}
     \label{fig:model_system}
     \end{center}
\end{figure}

Fig.~\ref{fig:model_system} {shows the starting model of our simulation}, consisting of a central SMBH of mass $M$ orbited by two separate stellar mass black holes, referred to as the inner and outer or (perturber) black holes. These are of mass $m_1$ and $m_2$ respectively and are initially placed in nearly circular, nearly coplanar orbits with semimajor axes $a_1$ and $a_2$ where $a_1 < a_2$. 

The separation of the stellar mass black holes (sBHs) is set in units of the mutual Hill radius (see~\cite{HAMILTON199243} for a derivation), which determines the separation between the sBHs where their orbits around the SMBH become unstable due to strong gravitational interactions between the two sBHs:

\begin{equation}\label{eq:mutual-hill-radius}
    R_{ H12} \equiv \frac{a_1+a_2}{2} \left(\frac{m_1+m_2}{3M}\right)^{1/3}
\end{equation}
where $m_1$ and $m_2$ are the masses of the stellar mass black holes, $a_1$ and $a_2$ are the semimajor axes of the black holes with respect to the SMBH, and $M$ is the mass of the SMBH. The separation when the orbits become unstable is typically around $R_{H12}$, depending on the gas frictional force acting on the two black holes \citep{Li2022}. Without the effect of gas, the two sBH orbits become unstable at a separation of $2\sqrt{3} R_{H12}$ \citep{Gladman93}. In our fiducial simulations, we start the sBHs at a separation of $5 R_{H12}$ and allow them to migrate close to each other due to gaseous drag.  

To determine whether the encounter of the sBHs are close enough and would emit significant GW, we calculate the separation in terms of the gravitational radii:
\begin{equation}\label{eq:grav_radius}
        R_{g, i} = \frac{Gm_i}{c^2}
\end{equation}
and we estimate the amount of GW radiation when the separation of the two sBHs is shorter than $50$ times the sum of their gravitational radii ($R_{g, 1} + R_{g, 2}$).
%\[R_g = \frac{G M}{c^2}\]

\subsection{Migration Forces}
To include gas effects in the AGN disk, we adopt two simple models following \citet{Li2022}, which represent the case of gas drag and migration traps separately. Specifically
the \textit{drag} force per unit mass ($\textbf{F}_{\text{drag}}$) can be expressed as the following:

\begin{equation}\label{eq:drag_force}
    \textbf{F}_{\text{drag}} = - \frac{\textbf{v} - \textbf{v}_{\text{K}}}{\tau_{\text{drag}}}
\end{equation}
where $\textbf{v}_{\text{K}}$ is the Keplerian velocity vector $\sqrt{GM/r^3} \,\hat\theta$, $r$ is the instantaneous distance between the SMBH and the \bh{} the force is applied to, $\textbf{v}$ is the \bh{} instantaneous velocity vector, and $\tau_{\text{drag}}$ is the characteristic timescale of the migration. 

For the case of migration {\em trap} at radius $r_0$ we implement a force per unit mass as the following,

\begin{equation}\label{eq:trap_force}
    \textbf{F}_{\text{trap}} = - \frac{\Omega_{K,0} (r - r_0) }{\tau_{\text{trap}}}\,\hat\theta
\end{equation}
where $\Omega_{K,0} = \sqrt{GM/r_0^3}$ is the Keplerian frequency, the trap radius $r_0$ is $a_1$, the initial semimajor axis of the inner \bh{}, and $\tau_{\text{trap}}$ is the characteristic timescale of this migration prescription. Throughout the text we use ``migration constant" to refer to both characteristic timescales $\tau_{\text{drag}}$ and $\tau_{\text{trap}}$. In our simulations the migration force is only applied to the secondary \bh{,} causing it to migrate inwards and interact closely with the inner \bh{.} This is because we are more interested in the difference of the migration rates between the stellar mass \bh{s} rather than {tracking the migration of} %migration strengths on the 
individual bodies. %This convergence in the separation of the stellar mass black holes also means it is redundant to vary their initial separations.% (see more details in section \ref{sec:parameter-space}). %in the study of the parameter space. 

\subsection{General Relativistic Prescriptions} \label{sec:gr-prescriptions}

Our first goal is to  establish whether different GR prescriptions significantly impact the results of the simulation. For this, we compare the occurrence of close encounters and their properties under no post-Newtonian (PN) effects, solely first order effects (1-PN), solely 2.5 order effects (2.5-PN), and simultaneous 1-PN and 2.5-PN effects. We use \rebound{}\footnote{\url{https://reboundx.readthedocs.io/en/latest/}} to include the first order PN effect, and we follow \citet{Blanchet06} to add the 2.5 order. Note that previous works often omitted 1-PN and included 2.5-PN only in their scattering experiments \citep[e.g.,][]{Samsing22}. Here, we demonstrate the importance of including both 1-PN and 2.5-PN effects.

For illustration we use a SMBH mass of $10^6 M_\odot$, equal mass inner and outer black holes of $20 M_\odot$, an initial inner black hole semimajor axis of $10^3$ $R_{g, {SMBH}}$ ($\sim9.87$~AU), and a semimajor axis separation of 5 $R_{H12}$ between the stellar mass black holes. The inclinations and eccentricities of the stellar mass black holes are sampled randomly for each simulation. For this portion, the inclination is sampled from a normal distribution with a mean of 0 radians and a standard deviation of 0.00087 radians (0.05 degrees). We include very low inclination here to allow more scatterings between the sBHs and to compare between the GR prescriptions more effectively. The eccentricities are sampled from a normal distribution with mean 0.05 and a standard deviation of 0.02. The mean anomalies are sampled from a uniform distribution from 0 to $2\pi$. 
%These initial parameters are collected in Table~\ref{tab:gr_test_params}. 
We implement the drag migration force from Eq.~\ref{eq:drag_force}. The migration constant varies from $10^{5.5}$ to $10^{6.5}$ times the initial period of the inner black holes around the SMBH. We count a passing of the stellar mass black holes as a ``close encounter" if they come within 100 $R_{g, 1}$ between each other. Note that we focus on the effects of different GR prescriptions here, and we explore the parameter space and study the dependence on the distribution of the orbital parameters and migration rates in detail in section \ref{sec:parameter-space}.

\begin{figure}[t]
\centering
     \includegraphics[width=\columnwidth]{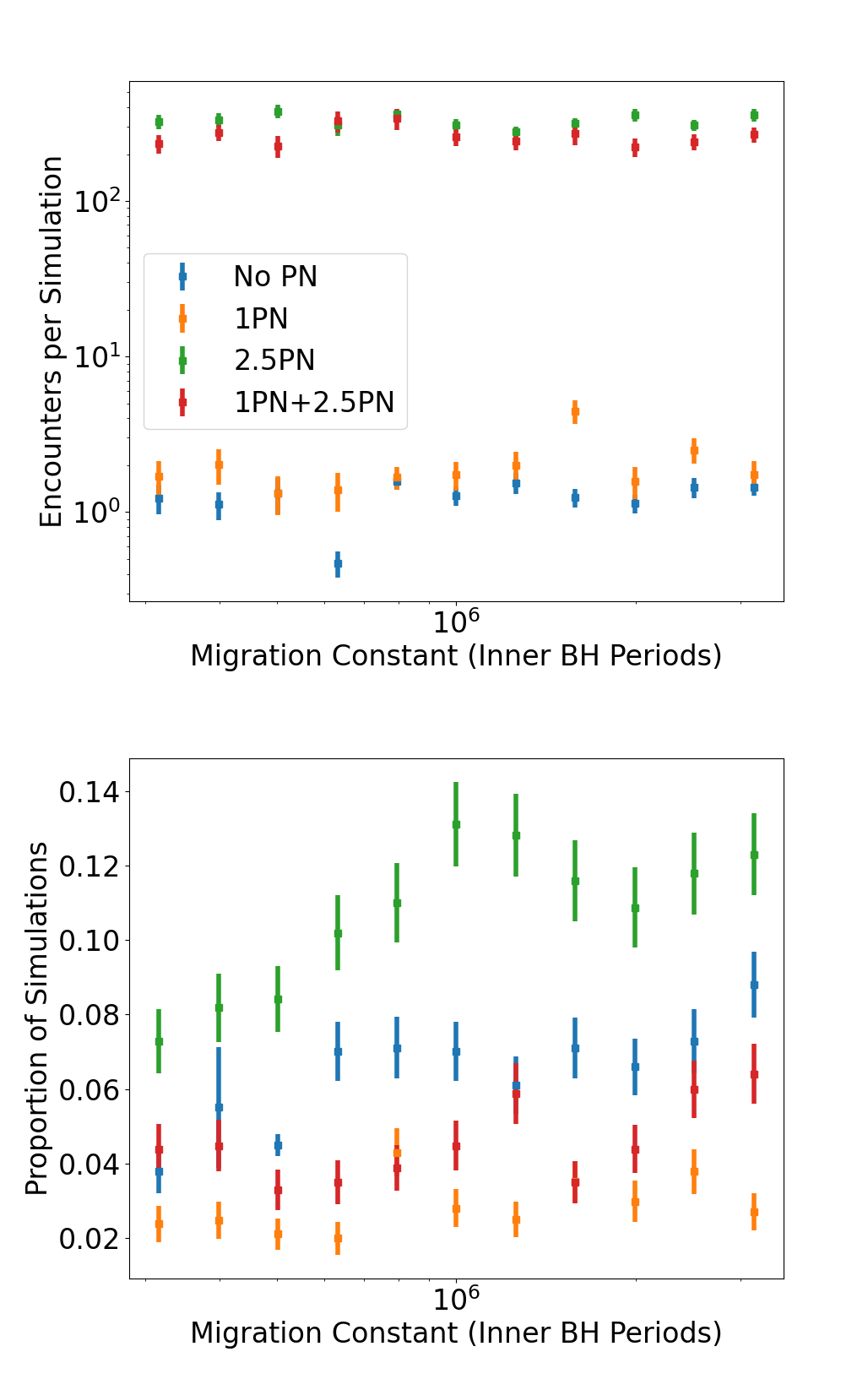}
     \caption{Comparison of the rate of sBH encounters depending on GR prescription and migration constant for the example simulation set described in \ref{sec:gr-prescriptions}. The top plot concerns the mean number of close encounters that occur in a simulation for a given migration constant. The bottom plot shows the proportion of simulations that experience any close encounters during the simulation time. The migration constant is for the drag prescription and is in units of the inner black holes' initial period around the SMBH. Each data point corresponds to 1,000 individual simulations and the uncertainty is from the Poisson uncertainty coming from the number of encounters and simulations. We find that including 1-PN, the encounter rate is suppressed comparing to the case including only the 2.5PN effect. }
     \label{fig:gr_prescrip_rate}
\end{figure}

Fig.~\ref{fig:gr_prescrip_rate} shows the rates of close encounters for simulations with different migration constants and GR prescriptions. Each point corresponds to a set of 1000 simulations with the same migration strength and GR prescription. The uncertainties included here are derived from the Poisson noise of the observed quantities. 

The sBHs can experience a large number of close encounters in a given simulation when their orbits become bound around each other. How does the number of encounters in each simulation vary depending on the GR prescription? The upper panel of Figure \ref{fig:gr_prescrip_rate} plots the number of encounters per simulation. In simulations that did experience encounters, the number of encounters were higher when 2.5-PN effects were included. Comparing the case with 2.5-PN only and the case with both 1-PN and 2.5-PN, the number of encounter in each individual simulation is suppressed slightly when both 1-PN and 2.5-PN are included.

The lower panel plots the fraction of simulations that have close encounters between the two sBHs. We find that simulations are less likely to have close encounters if the first-order effects are included (in comparison to the case without GR prescription). This is because orbital precession due to 1-PN can suppress close encounters between the sBHs within $\sim 100 R_g$. On the other hand, simulations with only 2.5-PN have higher likelihood to experience sBHs close encounters, due to the emission of GW energy. Comparing the case with 2.5-PN and the case with both 1-PN and 2.5-PN, the addition of 1-PN suppresses the number of systems that experience close encounters.

In summary, we find that the simulations including only 2.5-PN have higher close encounter rates compared to the cases with both 1-PN and 2.5-PN simulations. The differences are significant enough that we include both 1-PN and 2.5-PN in the rest of simulations used for analysis. Note that previous works including only 2.5-order PN may over-predict the rate of close encounters and GW emission \citep[e.g.,][]{Samsing22}.

%\filbreak

\section{Dependence on Parameter Space} \label{sec:parameter-space} 
To understand how the general parameters of the system affect the rate and quality of close encounters, we gradually expand the parameter space of our study. We do this through a series of simulation sets with differing initial parameters with each set consisting of 5000 independent runs.
These parameters are collected in Table~\ref{tab:varied_params}. 
For all of the simulations in this section, we sample the masses of each sBH from a uniform distribution of 10 to 150 $M_\odot$, in order to investigate how the encounters depend on the sBH masses and mass ratios. The maximum simulation length is $10^5$ periods of the initial period of the inner \bh{} around the SMBH, usually on the order of 3000-4000 years. The simulation can be ended earlier if the bodies collide (i.e. their separation distance is less than twice the sum of their event horizons $\sim 4 R_g$ for the equal mass sBH binaries assuming zero spins) or leave the system entirely. Note that only a small fraction of systems experience collisions ($\sim 0.68\%$ in the fiducial simulation set) or ejections ($\sim0.5\%$ in the fiducial simulation set). In the following, we detail the dependence of the rate of GW \bremss{} on sBH mutual inclination, migration rate, \& SMBH masses in section \ref{sec:parameter-space-dependence} and the dependence on sBH masses in section \ref{sec:mmr}.

\begin{table*}[htp]
\centering
\hspace*{-1.5cm}\begin{tabular}{|l|l|l|l|l|l|l|}
\hline\hline
Simulation Name & SMBH Mass                        & Initial $a_1$                       & $a_2 - a_1$                   & Inclination $(\mu,\sigma)$        & Migration             & Timescale               \\ \hline
Fiducial        & \multirow{12}{*}{$10^6 M_\odot$} & \multirow{17}{*}{$10^3 R_{g,SMBH}$} & \multirow{17}{*}{$5 R_{H12}$} & $(0, 0.05)$ rad                   & \multirow{7}{*}{Drag} & \multirow{3}{*}{$10^6$} \\ \cline{1-1} \cline{5-5}
IncLow          &                                  &                                     &                               & $(0, 0.01)$ rad                   &                       &                         \\ \cline{1-1} \cline{5-5}
IncHigh         &                                  &                                     &                               & $(0, 0.2)$ rad                    &                       &                         \\ \cline{1-1} \cline{5-5} \cline{7-7} 
DragLow1        &                                  &                                     &                               & \multirow{14}{*}{$(0, 0.05)$ rad} &                       & $10^5$                  \\ \cline{1-1} \cline{7-7} 
DragLow2        &                                  &                                     &                               &                                   &                       & $10^{5.5}$              \\ \cline{1-1} \cline{7-7} 
DragHigh1       &                                  &                                     &                               &                                   &                       & $10^{6.5}$              \\ \cline{1-1} \cline{7-7} 
DragHigh2       &                                  &                                     &                               &                                   &                       & $10^7$                  \\ \cline{1-1} \cline{6-7} 
TrapLow1        &                                  &                                     &                               &                                   & \multirow{5}{*}{Trap} & $10^5$                  \\ \cline{1-1} \cline{7-7} 
TrapLow2        &                                  &                                     &                               &                                   &                       & $10^{5.5}$              \\ \cline{1-1} \cline{7-7} 
TrapMed         &                                  &                                     &                               &                                   &                       & $10^6$                  \\ \cline{1-1} \cline{7-7} 
TrapHigh1       &                                  &                                     &                               &                                   &                       & $10^{6.5}$              \\ \cline{1-1} \cline{7-7} 
TrapHigh2       &                                  &                                     &                               &                                   &                       & $10^7$                  \\ \cline{1-1} \cline{2-2} \cline{6-7} 
LightSMBH1      & $10^5 M_\odot$                   &                                     &                               &                                   & \multirow{5}{*}{Drag} & \multirow{5}{*}{$10^6$} \\ \cline{1-1} \cline{2-2} 
LightSMBH2      & $10^{5.5} M_\odot$               &                                     &                               &                                   &                       &                         \\ \cline{1-1} \cline{2-2}
HeavySMBH1      & $10^{6.5} M_\odot$               &                                     &                               &                                   &                       &                         \\ \cline{1-1} \cline{2-2}
HeavySMBH2      & $10^{7} M_\odot$                 &                                     &                               &                                   &                       &                         \\ \cline{1-1} \cline{2-2}
HeavySMBH3      & $10^{8} M_\odot$                 &                                     &                               &                                   &                       &                         \\ \cline{1-1} \cline{2-2} \cline{3-3}
%FarAxis         & \multirow{2}{*}{$10^6 M_\odot$}  & $10^4 R_{g_SMBH}$                   &                               &                                   &                       &                         \\ \cline{1-1} \cline{3-4}
%FarSep          &                                  & $10^3 R_{g,SMBH}$                   & $10 R_{H12}$                  &                                   &                       &                         \\ 
\hline\hline
\end{tabular}
\caption{Parameters for the simulations sets with expanded parameter space. The migration constant is expressed in units of the initial period of the inner stellar mass black hole around the SMBH.}
\label{tab:varied_params}
\end{table*}

\subsection{Dependence on Inclination, Migration Rate and SMBH Masses}\label{sec:parameter-space-dependence}
First we consider how varying the standard deviation of the inclination distribution affects close encounters. We use the values of 0.01, 0.05, and 0.2 radians in the {\tt IncLow}, {\tt Fiducial}, and {\tt IncHigh} simulation sets respectively. A close encounter is marked when the distance between the stellar-mass black holes is within 50 times the sum of their gravitational radii. We monitor encounter rate and the GW emission energy $\Delta E_{GW}$ during each close encounter over the lifetime of the simulation. The emission energy can be calculated as follows:
\begin{equation}
    \Delta E_{GW} = \frac{85\pi}{12\sqrt{2}}\frac{G^{7/2} \mu^2 m_{12}^{5/2}}{c^5 r_{p}^{7/2}}
\label{delta e_gw}
\end{equation}
with $r_{p}$ the periastron distance of the binary, $m_{12}$ the combined mass of the sBHs ($m_1 + m_2)$, and $\mu$ the reduced mass $m_1 m_2 / m_{12}$ \citep{PhysRev.136.B1224, Quinlan1989}.

In each simulation, the sBHs can have a large number of close encounters (e.g., see Figure \ref{fig:gr_prescrip_rate}). We track maximum gravitational energy emitted in the series of encounters for each simulation, (i.e. the largest $\Delta E_{GW}$ released by a single close encounter during a simulation run), as it is a better indicator for the detectability of the encounter than the total $\Delta E_{GW}$ released over any number of encounters. We find that generally a single high-intensity encounter tends to dominate the simulation's cumulative $\Delta E_{GW}$ and conclude that cumulative \& peak $\Delta E_{GW}$ are closely correlated. We also check the percentage of simulations that emit a peak GW energy greater than 1 $M_\odot$ and 3 $M_\odot$ respectively, important thresholds in the detection of GW events by LIGO (assuming an observational distance of $\sim 1$ Mpc). 

The final eccentricities between the stellar mass black holes at the end of nearly every simulation is extremely close to unity, often within $\sim 10^{-10}$ for the encounters with significant GW emission, so we choose to not analyze any further dependence. 

\begin{figure} [t]
    \centering
    \includegraphics[width=\columnwidth]{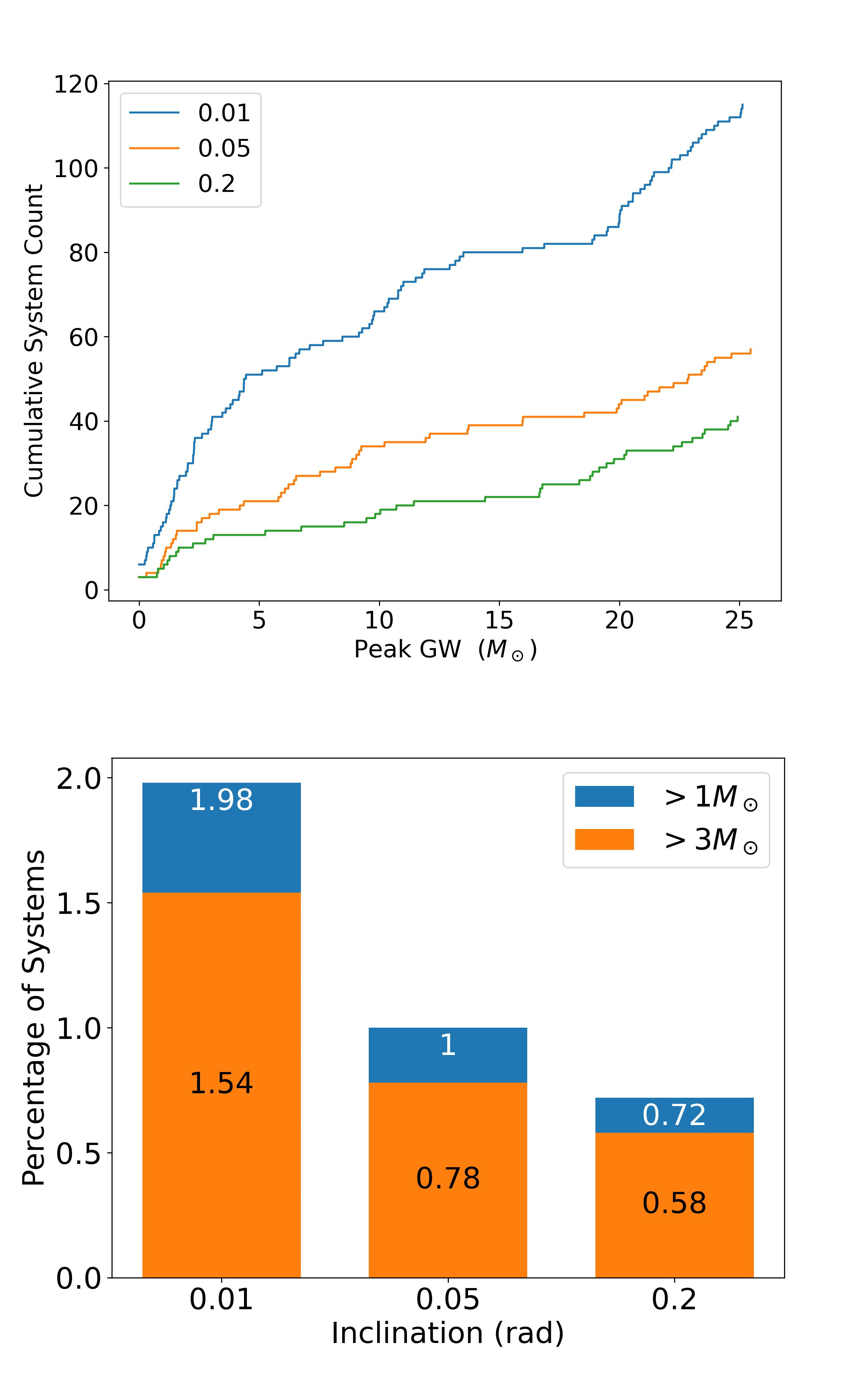}
    \caption{Analysis of close encounters in systems with varying standard deviation of inclination. The different runs IncLow, Fiducial, {and} IncHigh are labeled by the standard deviations of their inclination distributions expressed in radians. In the upper panel the y-axis shows the cumulative count of simulated systems over the peak $\Delta E_{GW}$ emitted over the entire simulation. The lower panel is the analysis of which simulation sets have more close encounters with peak $\Delta E_{GW}$ greater than 1 $M_\odot$ and 3 $M_\odot$. We see that lower inclinations lead to more intense encounters, as expected.}
    \label{fig:inc-merge}
\end{figure}

Fig.~\ref{fig:inc-merge} shows the cumulative frequency of simulations as a function of peak GW energy emitted in a close encounter for different inclination distribution between the sBHs. As one would expect, more coplanar configurations lead to more frequent and more intense close encounters as shown in Fig.~\ref{fig:inc-merge}. However, as the inclination increases further ($\gtrsim 3^\circ$), the decay in the encounter rate as inclination increases is small.

\begin{figure} [t]
    \centering
    \includegraphics[width=\columnwidth]{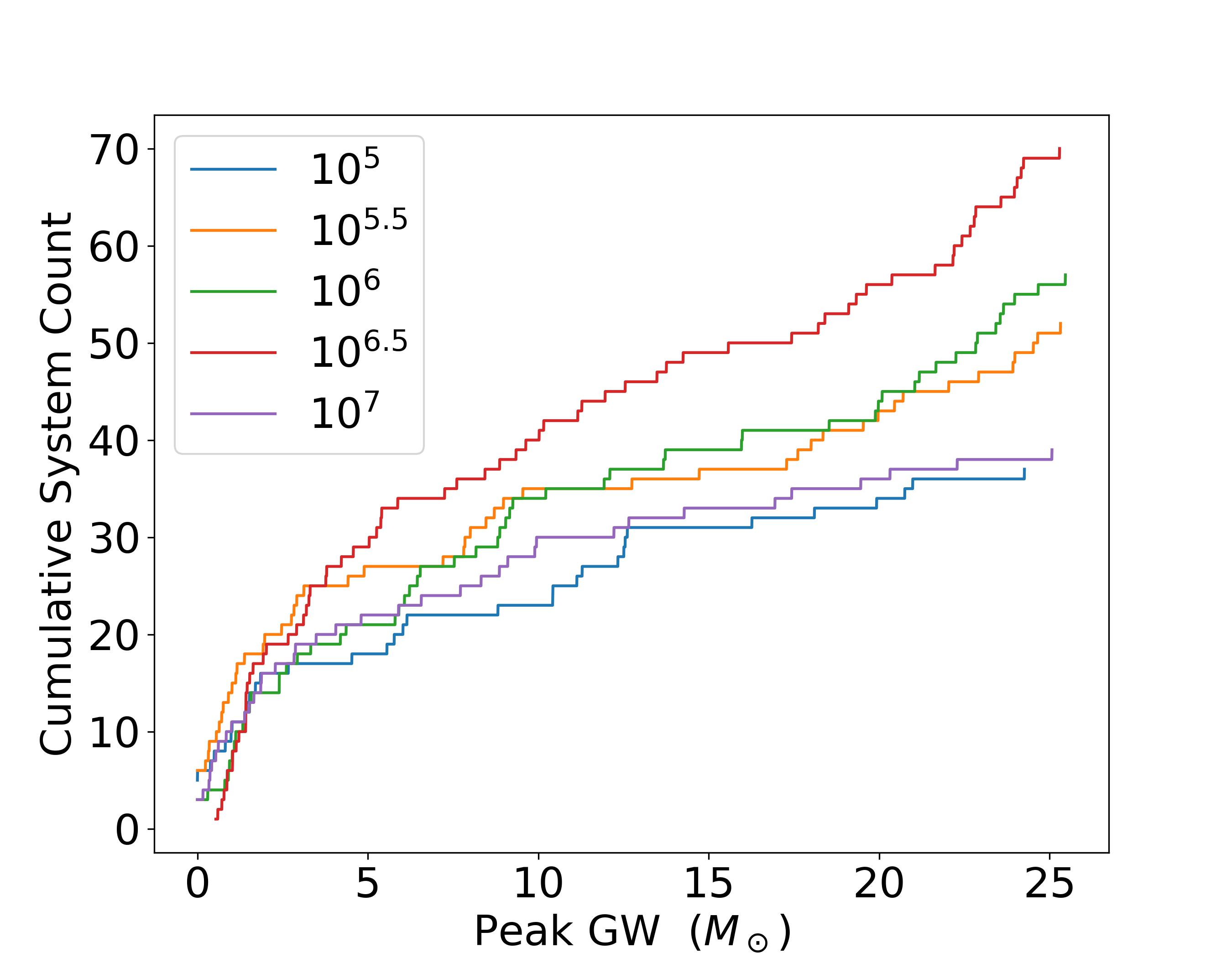}
    \includegraphics[width=\columnwidth]{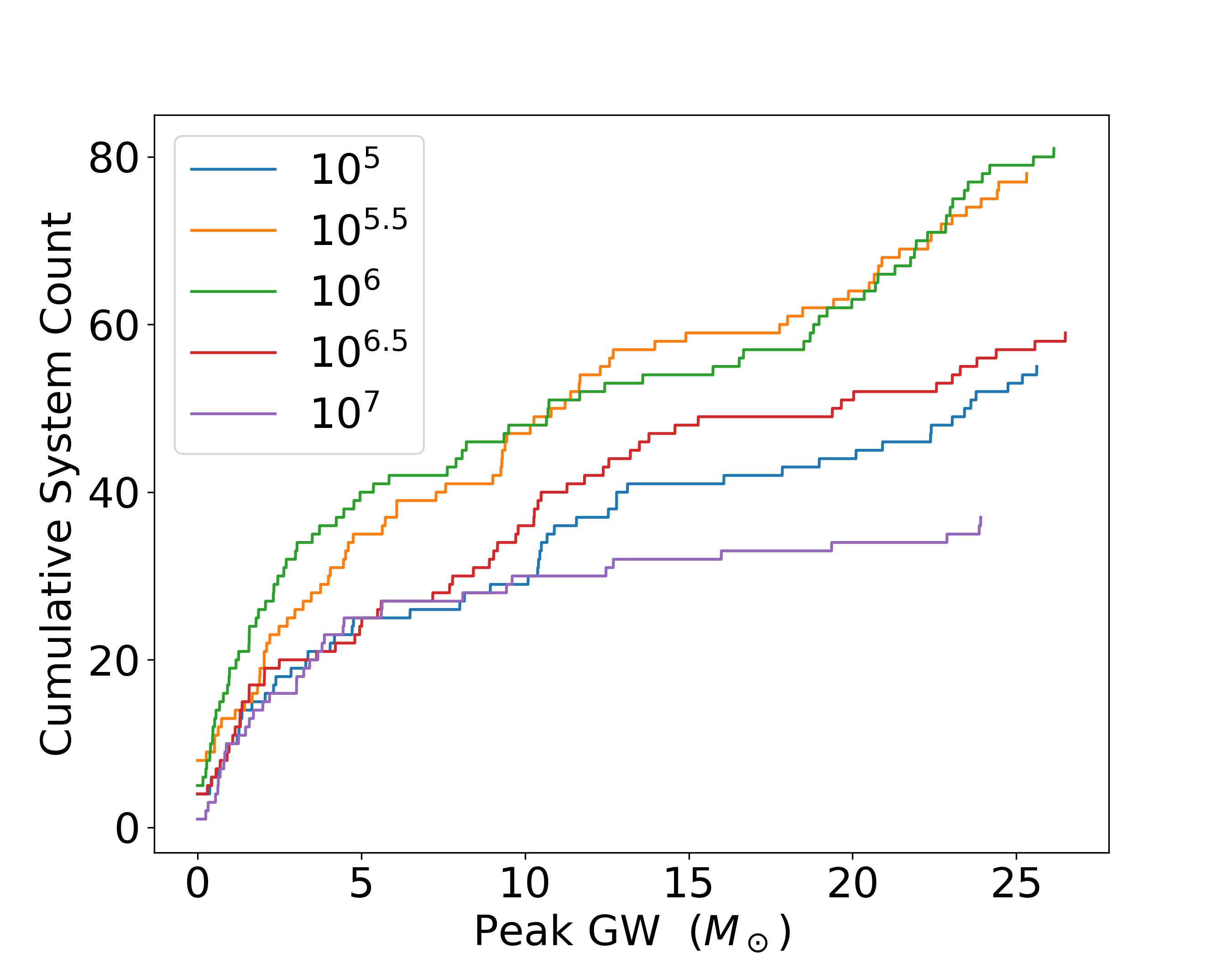}
    \caption{Analysis of close encounters in systems with varying strengths of the \emph{drag} (upper panel) and \textit{trap} (lower panel) migration force. The different runs (e.g. DragLow1, DragLow2, Fiducial, DragHigh1, \& DragHigh2) are labeled by the timescale of the migration (in initial periods of the inner black holes around the SMBH). The y-axis shows the cumulative count of simulated systems over the peak $\Delta E_{GW}$ emitted over the entire simulation. The most/strongest encounters occur in simulations with a migration timescale around $10^{5.5} - 10^{6.5}$ inner orbital period.}
    \label{fig:drag-merge}
\end{figure}

We then examine varying the migration constant in the drag force prescription from $10^5$ to $10^7$ times the initial period of the inner sBH around the SMBH, corresponding to simulation sets {\tt DragLow1}, {\tt DragLow2}, {\tt Fiducial}, {\tt DragHigh1}, and {\tt DragHigh2} respectively. Fig.~\ref{fig:drag-merge} demonstrates an enhancement in encounter rate around the range of $10^6$ to $10^{6.5}$ for the migration constant, matching our initial study of GR prescriptions. We note that we choose the range for the migration constant such that the migration is not too fast to prevent strong interactions between the two sBHs, but fast enough for the sBHs to approach each other within the simulation time. %Similar to the inclination study, these enhancements in occurrence match with enhancements in GW energy intensity, as shown in Fig.~\ref{fig:drag-merge}. 

We then instead implement the trap force from Eq.~\ref{eq:trap_force} with the same migration constant range as in the drag force runs. These correspond to the simulation sets {\tt TrapMed}, {\tt TrapLow1}, etc., but does \emph{not} include the Fiducial set. The enhancements in encounters rate \& intensity in these simulation sets indicate a slightly lower critical migration constant near $10^{5.5}$ to $10^6$, as seen in figure Fig.~\ref{fig:drag-merge}. Such a critical migration timescale exists because if migration is too strong, the stellar mass black holes will pass by each other too quickly to significantly interact. If migration is too weak, the black holes can enter mean motion resonance (discussed in Section~\ref{sec:mmr}) and will be unable to interact.

\begin{figure}[t]
    \centering
    \includegraphics[width=\columnwidth]{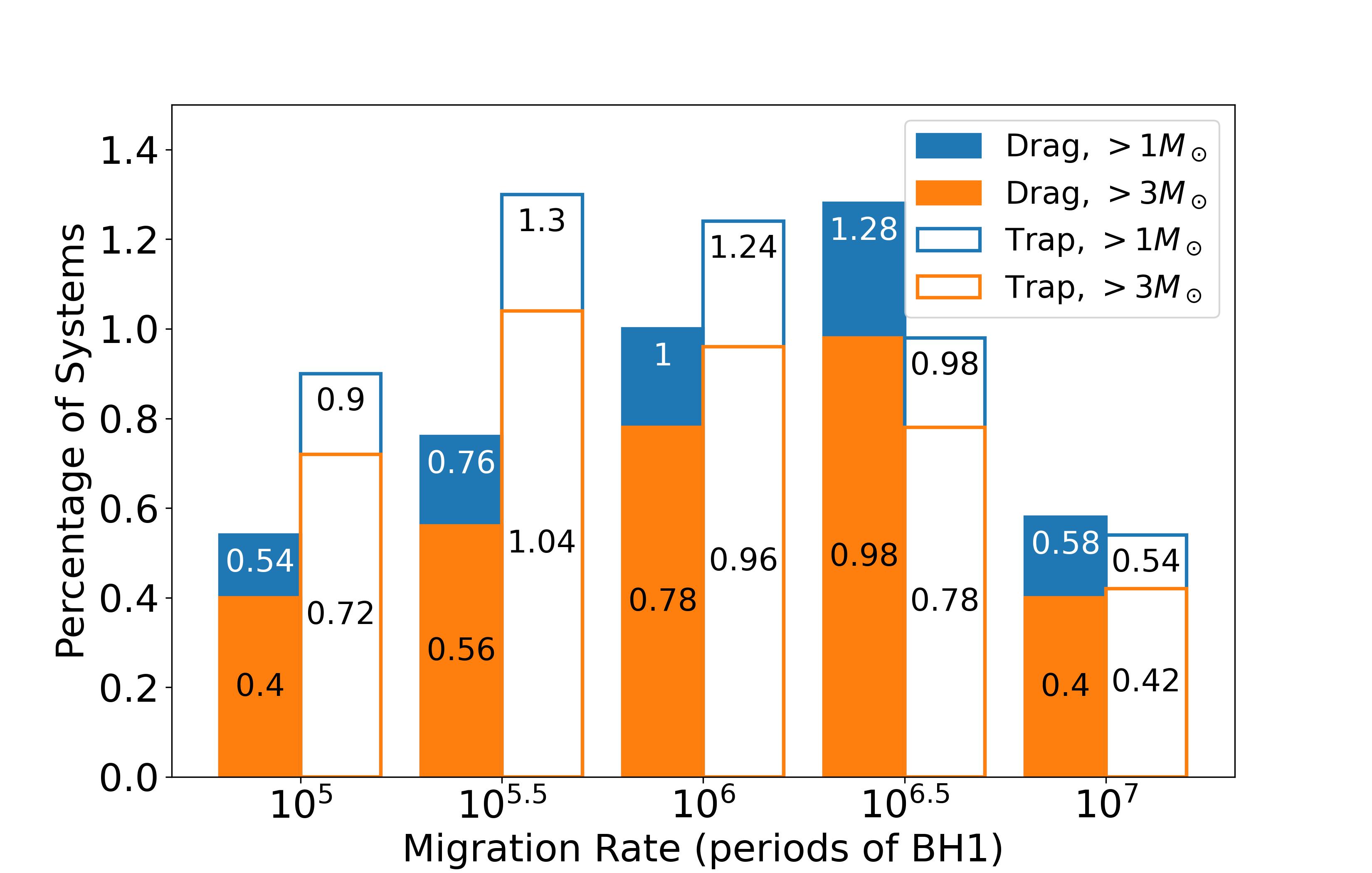}
    \caption{
    Analysis of which simulation sets have more close encounters with peak $\Delta E_{GW}$ greater than 1 $M_\odot$ and 3 $M_\odot$. The different runs are labeled by the timescale of the migration. In this plot we compare both migration prescriptions to each other. The migration rates with the most encounters here match those in Fig.~\ref{fig:drag-merge}.}
    \label{fig:gw-bar}
\end{figure}

In Fig.~\ref{fig:gw-bar} we compare the different migration force implementations directly. In general the trap force appears to have slightly greater than or equal rates of encounters with GW energy $>1 M_\odot$ and $>3 M_\odot$ when compared to the drag force implementation, except for the previously identified migration constant at $10^{6.5}$. Overall, the differences in the results due to drag v.s. trap forces are small.

We then study how the SMBH masses influences the results. We see a significant decrease in close encounter rates for higher mass central SMBHs. This is in part due to the fact that the initial semimajor axis of the inner black hole is calculated with the gravitational radius of the SMBH which scales with mass. This farther radial distance means any inclination results in larger absolute distances between the sBHs, reducing their interactions. We also have that the initial separation between the stellar mass black holes are correlated with the mass of the central SMBH. The initial separation of these black holes is proportional to the mutual hill radius $R_{H12}$. The initial semimajor axis of the inner black hole $a_1$ is determined as a multiple $A$ of the gravitational radius of the SMBH. We have that $a_2$ is very close to $a_1$, therefore

\begin{equation}
    R_{H12} \approx A \; \frac{GM}{c^2} \; \left(\frac{m_{12}}{3M} \right)^{1/3} \sim M^{2/3}
\end{equation}
% \[R_{H12} \approx A \cdot \frac{GM}{c^2} \cdot \left(\frac{m_{12}}{M} \right)^{1/3} \sim M^{2/3}\]

\begin{figure}[t]
    \centering
    \includegraphics[width=\columnwidth]{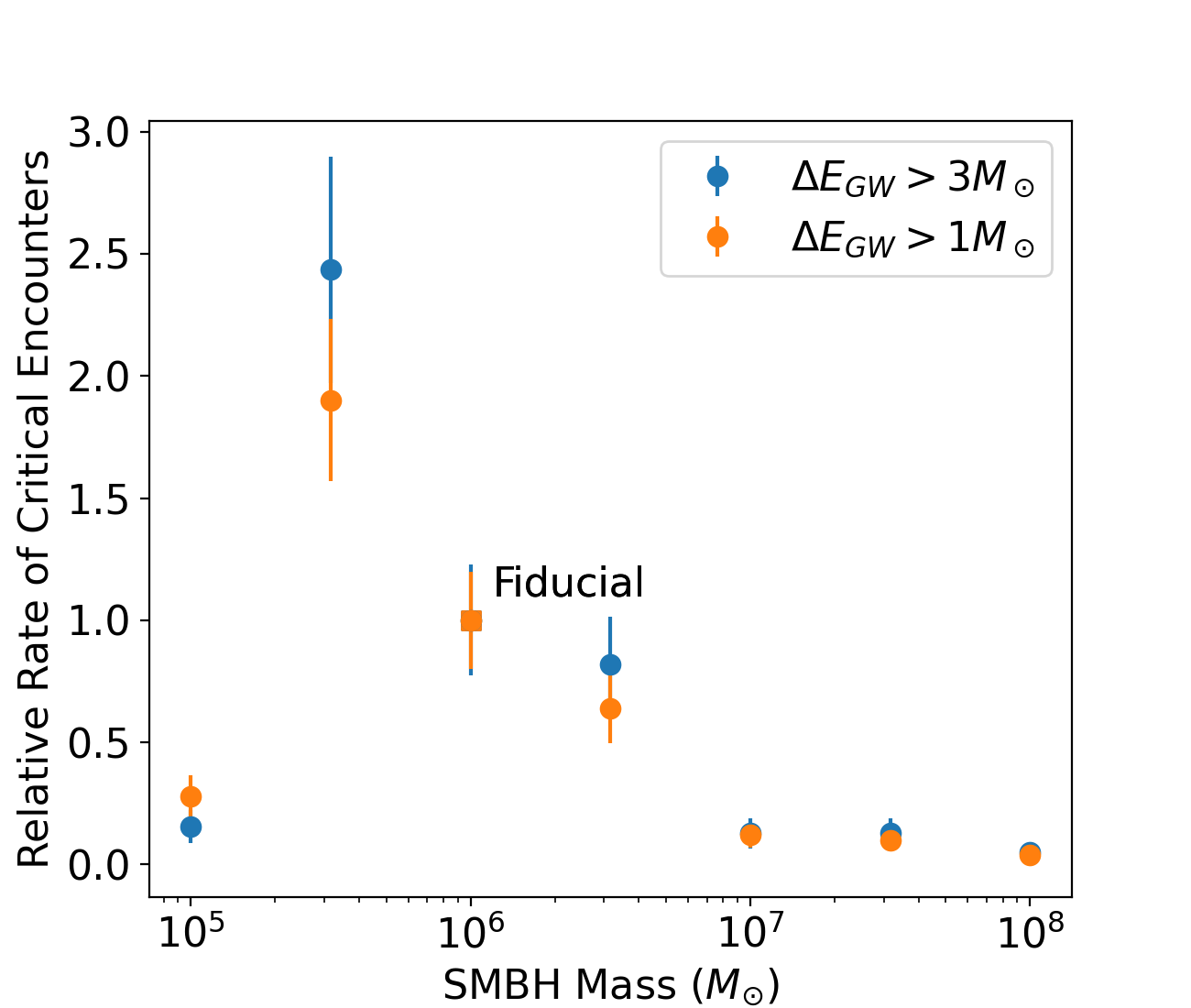}
    \caption{Comparison of the relative number of encounters that exceed critical values for $\Delta E_{GW}$  (see legend) for different masses of the central SMBH. These rates are expressed relative to the Fiducial run with a SMBH mass of $10^6 M_\odot$.}
    \label{fig:smbh-mass-rate}
\end{figure}

As the central mass grows so does the separation between the stellar mass black holes, reducing the encounter rate. This trend can be observed in Fig.~\ref{fig:smbh-mass-rate} which examines the rates of encounters with $\Delta E_{GW}$ that exceeds a critical threshold. The exception to this is for a particularly low mass SMBH (~$10^5 M_\odot$) we see a significant drop in the rate of these critical close encounters. We find that in this case the sBHs are more likely to become caught in mean motion resonance, preventing close encounters (further discussed in section~\ref{sec:mmr}). We also have that $R_{H12}$ becomes small enough such that it is more difficult for the sBHs to enter each other's spheres of gravitational influence and experience significant close encounters.

\begin{figure} [ht]
\centering
\includegraphics[width=\columnwidth]{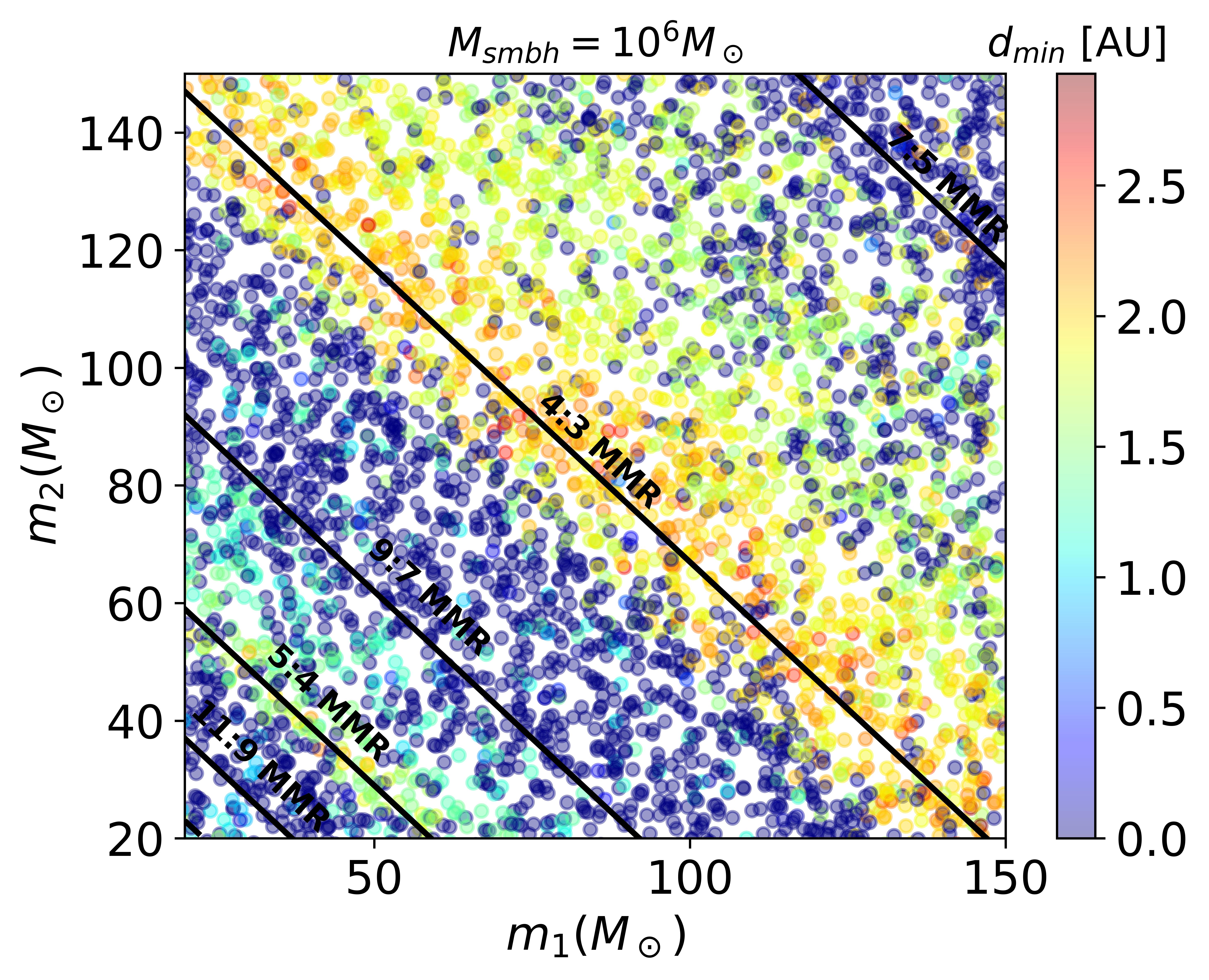}
\caption{Minimum distance attained in our ensemble of
	simulations as a function of mass of the inner and the perturber black hole masses. Color shows the minimum distance
	between the pair of black holes in our simulation. We can see that the outcome of the simulations depends on the total
	mass of the two black holes. Critical mass for capture into various MMR are shown as solid lines. \label{fig:m1m2dmin}}
\end{figure}

%\filbreak
\subsection{sBH-Mass Dependence, and Mean Motion Resonances}\label{sec:mmr}

In this section we investigate how the masses of the sBH influences the encounter rate and the GW emission. To illustrate the dependence on the sBH masses, we show in Figure \ref{fig:m1m2dmin} the results from our fiducial simulations (see Table \ref{tab:varied_params}). It shows the minimum distance between the two stellar mass black holes over the course of the simulation as a function of two parameters: the mass of the inner black hole (x-axis) and the mass of the perturber black hole (y-axis). We can see that the minimum distance depends on the masses of both of the black holes. The parameter space can be roughly divided into alternating regions where most black hole pairs either have close encounters ($d_{min}<0.1 $ AU) or they stay away from each other ($d_{min} \sim 1 $ AU). The boundaries between these regions depends on the total mass of the two black holes.

\begin{figure} [t]
\centering
\includegraphics[width=\columnwidth]{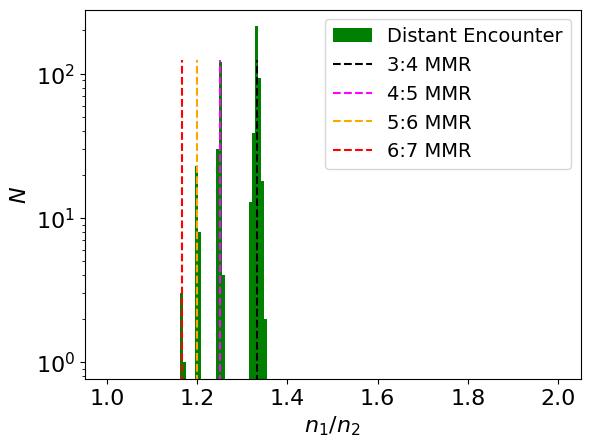}
\caption{Distribution of final period ratio for runs which did
not have close encounter (d\textsubscript{min }{\textgreater} 0.4 AU) in our ensemble of N-body simulations. The
location of relevant first order mean motion resonances are shown.}
\label{fig:prde}
\end{figure}

What leads to the alternating regions? Figure \ref{fig:prde}\ shows the histogram of the ratio of periods of the inner and the perturber black holes from our ensemble of simulations. We only include runs in which the black holes do not have close encounters ($d_{min} > 0.5$ AU). Also, the period ratios shown in the figure are calculated at the end of the simulation ($10^{5}$ periods of the inner black hole's orbit). We can see that most of the black holes have a period ratio of $\sim$ 1.33, 1.25, 1.2 and 1.16 corresponding to first order MMRs 4:3, 5:4, 6:5 and 7:6. Thus, Figure \ref{fig:prde} shows that most black holes pairs with $d_{min}>0.5$ AU are captured into mean motion resonances (MMRs). 

We now derive an analytical expression for the minimum total mass required for a black hole pair to be captured into MMR. If a black hole pair is captured into a $p:q$ MMR, the semi-major axis of the outer black hole is given by $a_{2,res} = a_1 f^{2/3}$, where $f=q/p$ is the ratio of the periods of the two black holes. Also, we initialize the semi-major of the outer black hole at 
$$ a_{2,hill} = a_1
\frac{1+\frac{k_{hill}}{2} \left(\frac{M_{tot}}{3M_{smbh}}\right)^{1/3} }{ 1- \frac{k_{hill}}{2} 
\left(\frac{M_{tot}}{3M_{smbh}}\right)^{1/3}  },$$  where $k_{hill}(=5)$ is the initial orbital separation in mutual hill radii. For the pair to be captured in a MMR we need, $ a_{2,hill} \geq a_{2,res}$. Equating the two expressions, we get the critical total mass: 

\begin{equation}\label{eq:mmrmtoteq}
M_{tot} = \frac{24 \left(f^{2/3}-1\right)^3 M_{smbh}}{\left(f^{2/3}+1\right)^3 k_{hill}^3}
\end{equation}

For example, if a pair of black holes orbiting a SMBH of mass $10^{6}$ $M_{\odot}$ are to be captured into a $4:3$ MMR ($f =4/3 \sim 1.33$) with  $k_{hill} =5$, they should have a minimum total mass of $M_{tot}$ = 167.76 $M_{\odot}$ (agreeable with Figure \ref{fig:m1m2dmin}). We can see that for a given total mass of the black hole pair $M_{tot}$, we increase the value of $k_{hill}$ to allow capture into wider range of MMRs. Increasing the mass of the SMBH to $10^{8}$ $ M_{\odot}$ would make capture into these MMRs impossible for $M_{tot} < 300 M_{\odot}$.

Critical total masses for various MMRs are shown as solid lines in Figure \ref{fig:m1m2dmin}. We can see that these lines mark the boundary between the alternating regions of the parameter space very well. We can also see that when the total mass of the black hole pair is such that they are captured in a first order MMR, the pair do not have a close encounter. On the other hand, when the pair is captured in a second order MMR, they are more likely to have a close encounter ending in a merger. Note \citet{Secunda2019} also found some very high-order MMRs around the migration trap.

\begin{figure}
\centering
\includegraphics[width=\columnwidth]{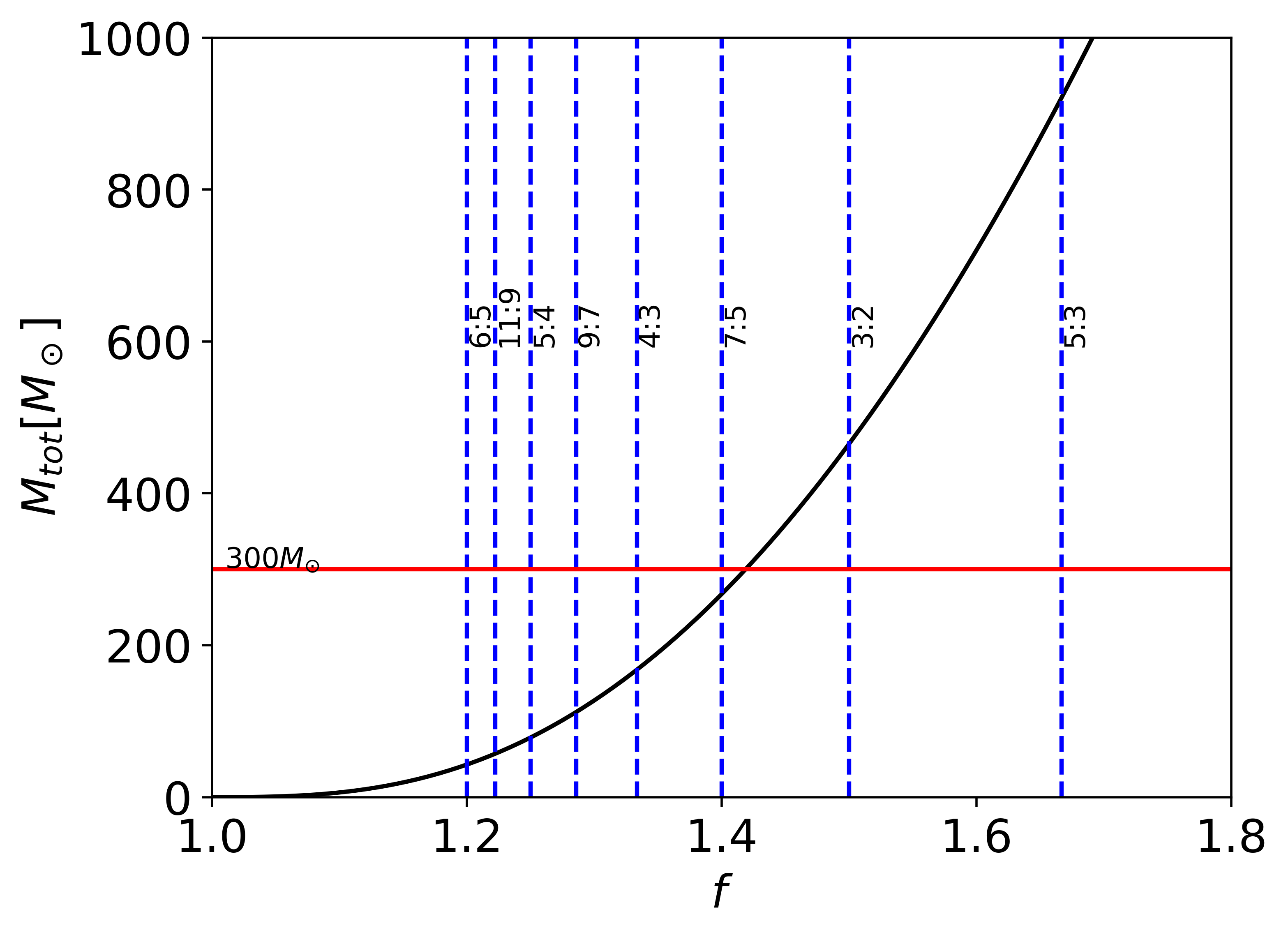}
\caption{\label{fig:mvf} $M_{tot}$ vs $f$ relationship as given by Equation \ref{eq:mmrmtoteq}. We take $M\textsubscript{smbh} = 10^{6} M_{\odot}$ and $k_{hill} = 5$. The nominal location of various MMRs is shown using the vertical dashed lines. The total mass of $300 M_{\odot}$ is shown using the horizontal line. We can see that for $M_{tot} < 300 M_{\odot}$ black holes can only be captured into a few MMRs.}
\end{figure}
Figure \ref{fig:mvf}\ shows $M_{tot}$ as a function of $f$ as given by Equation \ref{eq:mmrmtoteq}. It should noted that we are only interested in values of $f$ which correspond to ratios of small integers. This is because capture is possible only for low-order MMRs. Some of the resonances are marked in the figure as vertical lines. For $M_{tot} < 300 M_{\odot}$, we can see that capture into 3:4, 4:5, 5:6 and 6:7 first order MMRs is possible. In addition, capture into second order resonances 9:7, 11:9 and 7:5 MMRs is also possible. This is in agreement with our simulations in which most black holes are captured into 4:3 and 5:4 resonances. 

\begin{figure*}
\centering
\includegraphics[scale=1]{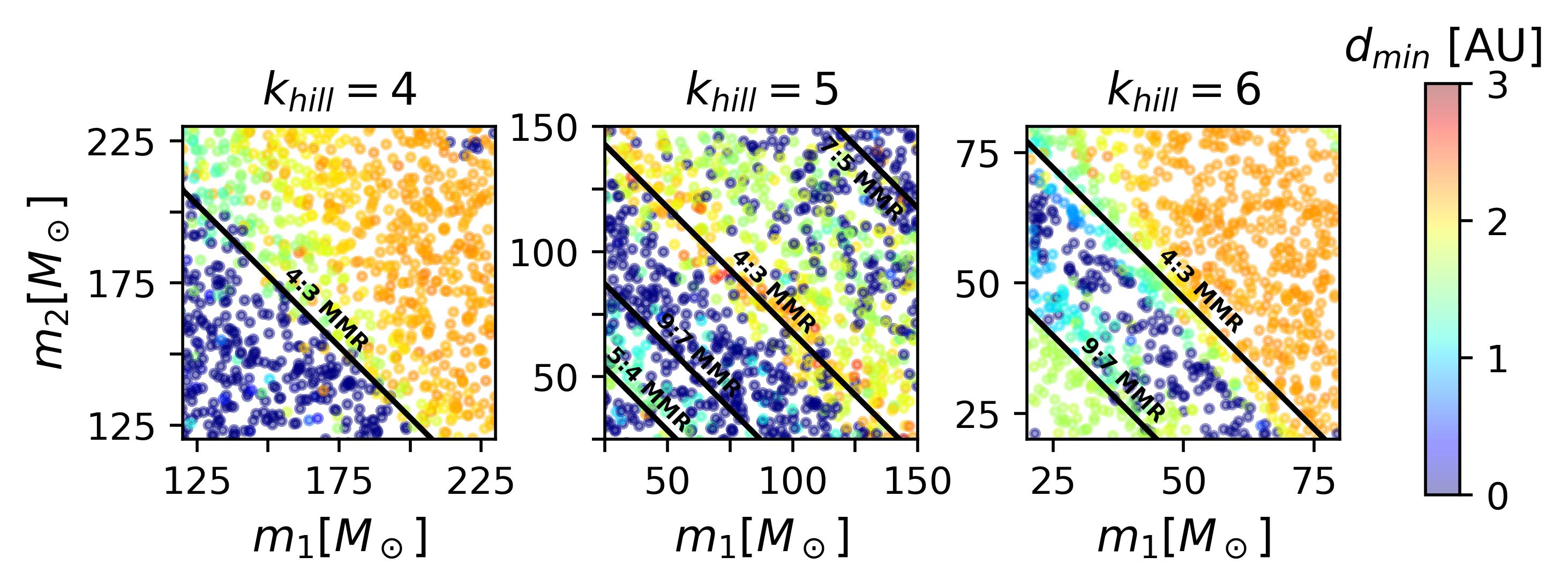}
\caption{\label{fig:m1m2khill} Resonance capture for different initial separations between the sBHs. The minimum separation between the sBHs is shown using the colors. The mass of the inner sBH is shown on the x axis, and the mass of the perturber sBH is shown on the y-axis. The minimum total mass needed for the sBH pair to be captured in a MMR is shown using the black lines (Eqn. \ref{eq:mmrmtoteq}). Between the panels we change the initial separation between the sBHs.}
\end{figure*}

It should be noted that our results depend sensitively on the choice of the initial separation between the sBHs.  Hence, we ran additional simulations in which we varied the initial separation between the sBHs. Our results are shown in Figure \ref{fig:m1m2khill}. In the left and the right panels we chose $k_{hill}=4$ and $6$ respectively. For comparison, the middle panel shows results from our fiducial runs. The alternating regions, whose boundaries are given by Eqn \ref{eq:mmrmtoteq}, can be seen in all the three panels. Also, smaller mass sBHs can be captured into MMRs if we increase the initial separation between them. Whether a sBH pair of a given total mass has a close encounter depends on it’s initial separation. For instance, when $M_{tot}=100 M_\odot$ blackholes are likely to have a close encounter if $k_{hill}=5$, and stay distant if $k_{hill}=6$.

In summary, we find that the encounters between the inner and the perturber black holes can be classified into three categories: 1) the pair can have strong interactions and emit significant GW during close scattering; 2) the perturber could be captured in a mean motion resonance with the inner black hole.  When this happens, both the inner and the perturber black holes migrate in a way which keeps their period ratio roughly constant. Consequently, black holes locked in MMRs cannot have close encounters, and hence cannot merge; 3) the black holes can have a minor interaction in which the migrating perturber briefly perturbs the orbit of the inner black hole. A brief jump in the semi-major axis of the inner black hole occurs as the perturber crosses the orbit of the inner black hole. The perturber black hole continues to migrate inwards, with the other black hole now on an wider orbit. Mergers are not possible during such encounters. We note that we neglected turbulences and additional orbiters (e.g., sBHs and stars) in the AGN disk, which may kick the sBHs out of the MMRs and increase the encounter rates.

%\filbreak
\section{Occurrence Rate of Bremsstrahlung}\label{sec:pop-synth}

In this section, we run a much larger set of simulations to explore the detection rate of the \bremss{} emission and how it compares with the rate of BH mergers. The estimated rates are sensitive to a wide number of parameters, such as the number of sBHs in the galactic nuclei, the AGN disk profile, the AGN lifetime, and the migration timescale. These numbers are highly uncertain. Thus, instead of calculating a definite rate using a complex disk model, we focus on simplified configuration with fiducial parameters. Our calculated rate can then be scaled to obtain the \bremss{} rate for specific systems with different AGN disk and sBH properties, and to incorporate the uncertainties to obtain an overall rate per volume. In the end, we discuss how our simulation parameters affect our rate estimate.

Our suite of simulations consists of 80,000 runs of sBH-sBH encounters. 
The initial inclinations \& eccentricities are sampled in the same way as the {\tt Fiducial} simulation set, while the masses of the inner and perturber black holes are sampled from a power-law initial mass function (IMF) with $dN/dm \propto m^{-2.36}$ following \citet[e.g.,][]{Tagawa20} instead of a uniform distribution as in section~\ref{sec:parameter-space}. 
These simulations use a single migration coefficient of $10^6$ periods of the inner \bh{} over the entire set, implementing the drag \& trap force respectively (40,000 simulations for each), and we set the central SMBH to be $10^6 M_\odot$. 
We choose a migration rate of $10^6$ and a central mass of $10^6 M_\odot$ in order to maximize the rate of close encounters in the simulations and save computational cost (following the results in section \ref{sec:parameter-space}), since only a small fraction of systems encounter scattering close enough to produce strong GW wave emission. 
We incorporate the range of SMBH masses in the rate calculation, 
and we discuss the effects of different migration rate in the end of the section.

\begin{figure}
    \centering
    \includegraphics[width=\columnwidth]{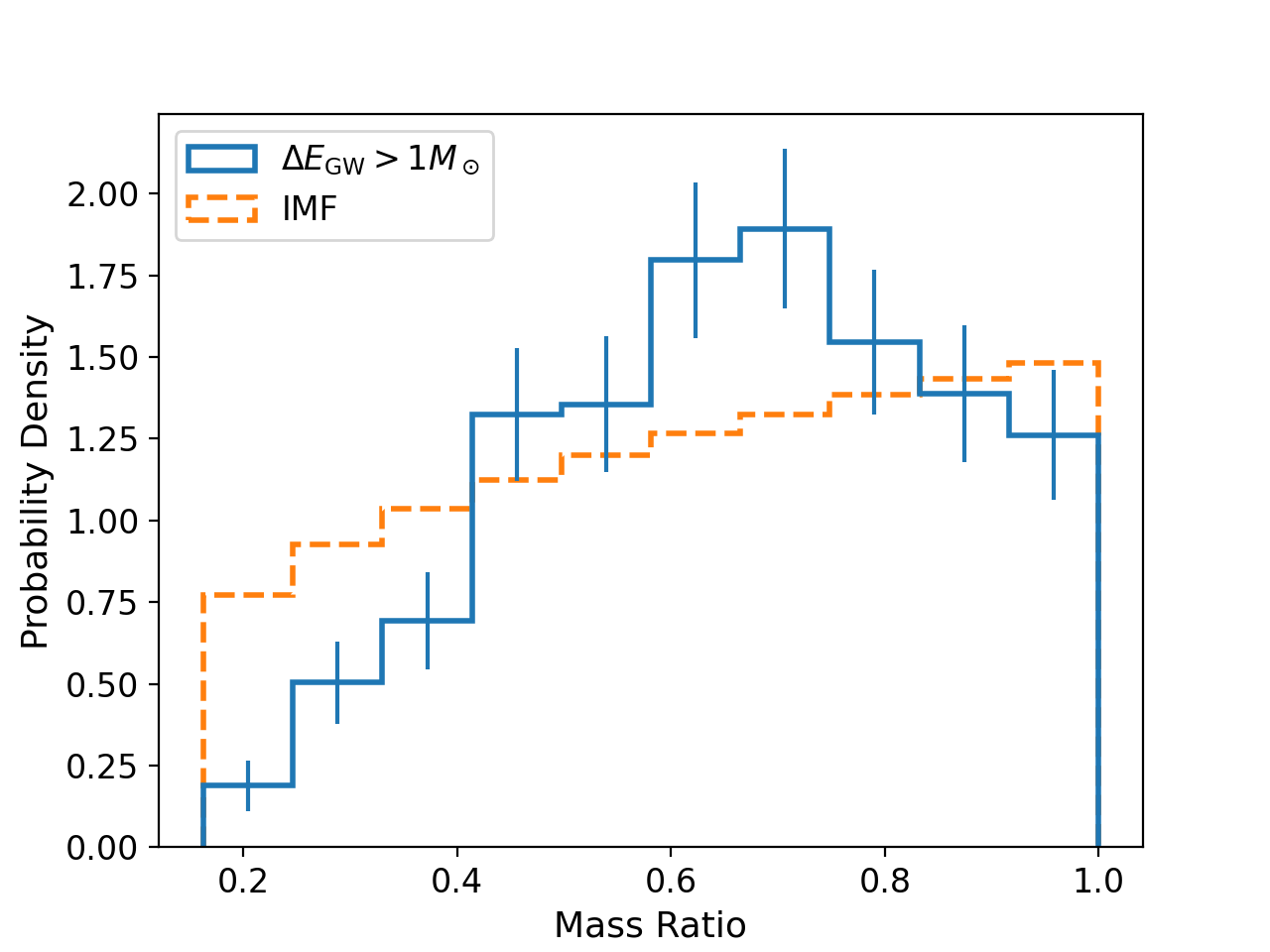}
    \caption{Probability density function of the mass ratio $q$ of systems that emitted a peak GW energy above $1 M_\odot$.}
    \label{fig:mass-ratio-1M}
\end{figure}

\begin{figure}
    \centering
    \includegraphics[width=\columnwidth]{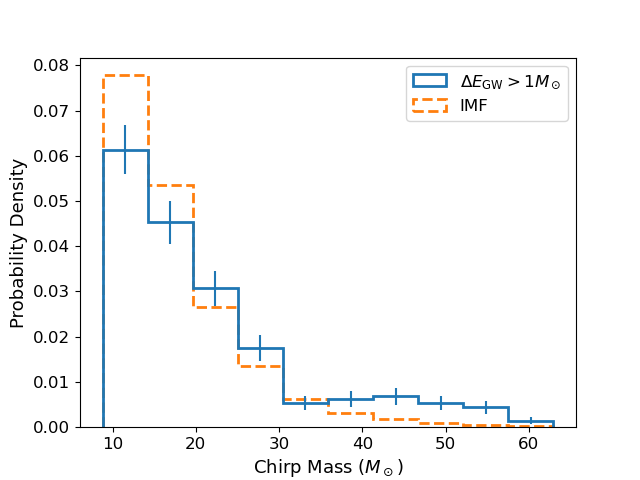}
    \caption{The probability density function of the chirp mass $\mathcal{M}_c$ of systems that emitted a peak GW energy above $1 M_\odot$.}
    \label{fig:chirp-ratio-1M}
\end{figure}

\begin{figure}
    \centering
    \includegraphics[width=\columnwidth]{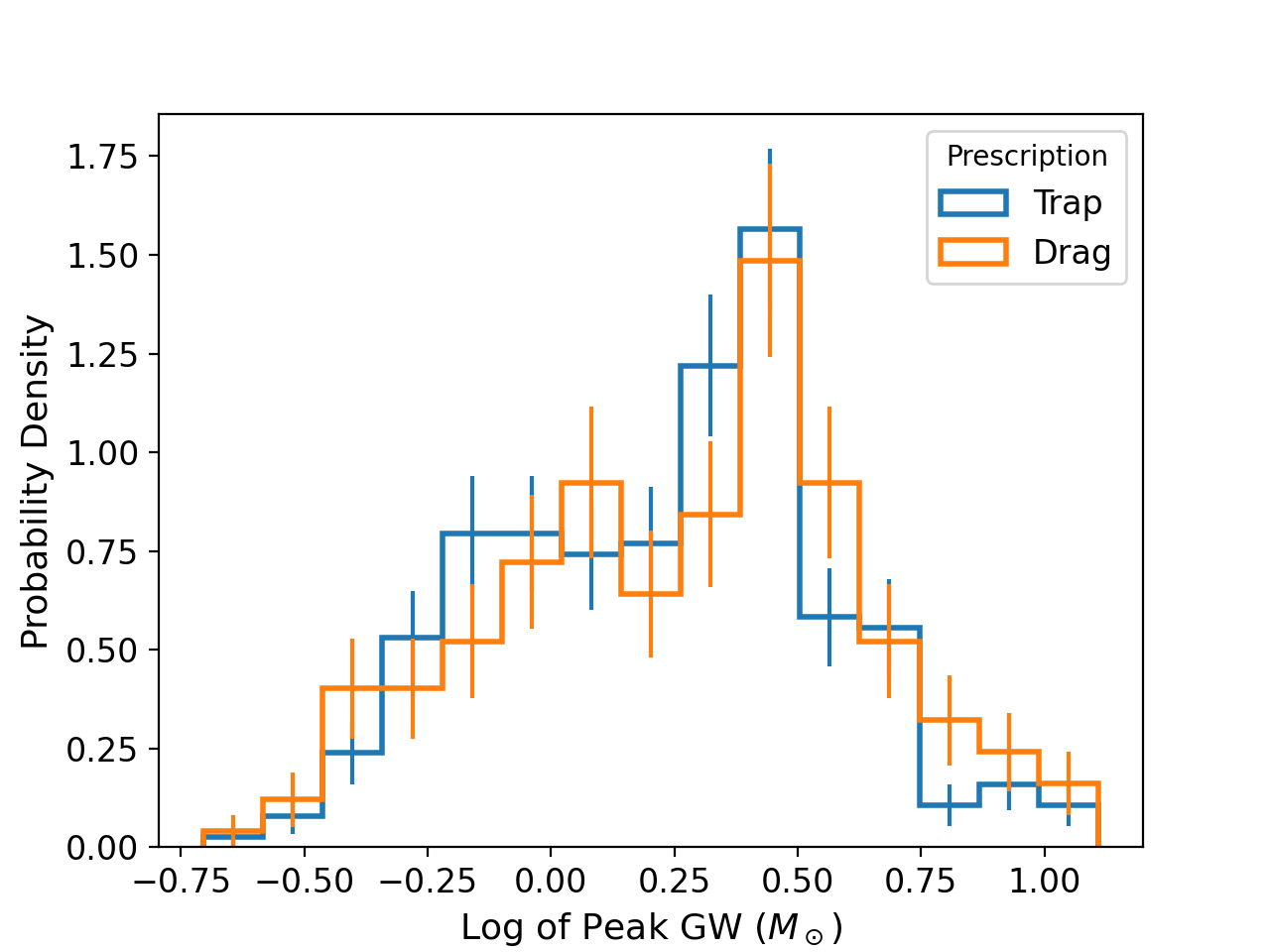}
    \caption{Probability density function of the peak GW energies over $0.1 M_\odot$ emitted during simulations with the Trap prescription (blue) and the Drag prescription (orange).}
    \label{fig:pop-synth-pdf}
\end{figure}

We use \rebound{} to run the simulations, and the detailed set up of the simulations can be found in section \ref{sec:gr-prescriptions}. Figures \ref{fig:mass-ratio-1M} \& \ref{fig:chirp-ratio-1M} show the distributions of mass ratio and chirp mass  of systems emitting at least $1 M_\odot$ of GW emission; from these, we can draw conclusions about the prominence of this merger channel for producing observable GW properties. Figure \ref{fig:mass-ratio-1M} shows that the mass ratio of systems that emit $> 1 M_\odot$ of peak GW power peaks roughly at $q \sim 0.7$ with a lower probability density tail towards lower values (a factor of $\sim 3$ less) than characteristic in the initial mass function the population was drawn from (dashed orange line in Figure \ref{fig:mass-ratio-1M}). 
Figure \ref{fig:chirp-ratio-1M}
shows that those same systems ($\Delta E_{GW} >1M_\odot$) had a chirp mass distribution weighted towards larger values comparing to the IMF (a $\sim 20\%$ supress in the lower mass bin of $\sim 10 M_\odot$), and a higher probability (a factor of $\sim 5$ comparing to the IMF) in the range of $> 35 M_\odot$. Given a total of $\sim 300$ systems that emit a peak GW energy above 1 $M_\odot$,the results in Figure \ref{fig:mass-ratio-1M} and \ref{fig:chirp-ratio-1M} are statistically significant (see the uncertainties). 

Fig.~\ref{fig:pop-synth-pdf} 
shows the probability density function of the peak GW energy emitted in the simulation. The simulations with the drag force implementation generally have slightly lower rate of strong peak GW emission than the trap force implementation when the migration is fast, following a similar trend as in Fig.~\ref{fig:gw-bar}. This is because sBHs are more likely to move close to each other following the ``Trap'' prescription, where the two sBHs migrate to the same distance from the SMBH at the ``Trap'' location. On the other hand, the sBHs can pass each other and miss close encounters more easily following a ``Drag'' prescription with faster migration rate. 
In all, $0.39\%$ of systems with the drag migration prescription emitted $> 1 M_\odot$ of GW energy, with that being $0.56\%$ with the trap prescription. For simulations that emit $>3 M_\odot$ of GW energy these are $0.145\%$ for the drag prescription and $0.19\%$ for the trap prescription. It shows that only a very small fraction of systems have strong GW emission during the scattering of the sBHs. We can next use these numbers to obtain the expected rate of detectable GW \bremss, scaling with SMBH masses, number of sBHs expected in the AGN disk, as well as the AGN disk number density and lifetime. 

To estimate the occurrence rate of systems emitting GW radiation above the observational threshold, we follow the calculation from \cite{Tagawa20} which provides an occurrence rate for stellar mass black hole mergers in AGN disks given merger fractions and typical system parameters,

\begin{equation}
    \mathcal R_{\text{sBH}} = \int_{M_{\text{SMBH,min}}}^{M_{\text{SMBH,max}}} \frac{d n_{\text{AGN}}}{d M_{\text{SMBH}}} \frac{f_{\text{BH,enc}} N_{\text{BH,cross}}}{t_{\text{AGN}}} dM_{\text{SMBH}}
    \label{eq:tagawa-73}
\end{equation}
where $f_{\text{BH,enc}}$ is the close encounter fraction per black hole that leads to detectable \bremss{} emission, $N_{\text{BH,cross}}$ is the number of black holes crossing AGN disks, $t_{\text{AGN}}$ is the average lifetime of AGN disks, $n_{\text{AGN}}$ is the average number density of AGNs in the universe, and $M_{\text{SMBH,min}}$ and $M_{\text{SMBH,max}}$ are the minimum and maximum SMBH masses. This integral can be expressed as

\begin{equation}
\begin{split}
    \mathcal R_{sBH} \sim & 3 \text{ Gpc}^{-3} \text{ 
     yr}^{-1} \left(\frac{f_{\text{BH,enc}}}{0.5} \right) \\
   & \left(\frac{t_{\text{AGN}}}{30 \text{ Myr}} \right)^{-1} \left(\frac{r_{\text{AGN,MW}}}{0.1 \text{ pc}} \right) \left(\frac{\eta_{\text{n,BH}}}{0.005 M_\odot^{-1}} \right)
\end{split}
    \label{eq:tagawa-81}
\end{equation}
over the interval of 
$M_{\text{SMBH}} = 10^6 - 10^8 M_\odot$, 
%$M_{\text{SMBH,min}} = 10^6, \ M_{\text{SMBH,max}} = 10^8$ 
where $r_{\text{AGN,MW}}$ is the typical size of the AGN disk for $M_{\text{SMBH}} = 4 \times 10^6 M_\odot$ and $\eta_{\text{n,BH}}$ is the number of black holes per unit stellar mass. 

We can use the simulation results ($f_{\rm brem,BH}$) to calculate the percentages of systems that experience critical close encounters, to derive $f_{\text{BH,enc}}$.
Different from $f_{\text{BH,enc}}$ (the encounter rate of \textit{sBHs in the AGN disk} to produce GW \bremss), $f_{\rm brem,BH}$ determines the probability that \textit{a pair of scattering sBHs} to produce GW \bremss. Thus, to calculate $f_{\text{BH,enc}}$, we must scale $f_{\rm brem,BH}$ by the number of scattering sBHs pairs in the AGN disk lifetime.
The number of black holes in a nuclear star cluster ($N_{\text{BH,NSC}}$) can be calculated as 

\begin{equation}
    N_{\text{BH,NSC}} = \eta_{\text{n,BH}} M_{\text{NSC}} \,
    \label{eq:tagawa-76}
\end{equation}
where $M_{\text{NSC}}$ is the mass of the nuclear star cluster with $\eta_{\text{n,BH}} \sim 0.005$ $M_\odot^{-1}$ and $M_{\text{NSC}} \sim 4 \times 10^6 M_\odot$ for typical nuclear cluster, following \cite{Tagawa20}. This gives $N_{\text{BH,NSC}} \sim 2 \times 10^4$. We assume that around 10\% of these black holes are actually in the AGN disk and 50\% interact with each other due to migration forces, since the migration timescale is short comparing to the AGN disk lifetime. Thus, each inner black hole will have $\sim 10^3$ encounters. Then, we scale the results by a factor of $10^3$ to get $f_{\text{BH,enc}} \sim 10^3 f_{\rm brem,BH}$. 

Because our simulations set $M_{\text{SMBH}} = 10^6 M_\odot$, we must also adjust these fractions for different SMBH masses. From \cite{Tagawa20}, the relative contributions to the rate calculation from each range of SMBH mass are 0.96\% for $10^5 - 10^6 M_\odot$, 34\% for $10^6 - 10^7 M_\odot$, 59\% $10^7 - 10^8 M_\odot$, and 6.1\% $10^8 - 10^9 M_\odot$. We consider the range from $10^6 - 10^8 M_\odot$ as these have the most significant contributions and match our parameter space exploration in section~\ref{sec:parameter-space}. We estimate the occurrence rates in these ranges using the relative rates observed in section \ref{sec:parameter-space-dependence} and shown in Fig.~\ref{fig:smbh-mass-rate}. 

In summary, assuming a $t_{\text{AGN}}$ of 30 Myr and an $r_{\text{AGN,MW}}$ of 0.1 pc, we obtain an occurrence rate of 3.2 $\text{Gpc}^{-3} \text{ yr}^{-1}$ for events with GW energy $>1 M_\odot$. The results for different migration prescriptions and GW threshold are shown in Table \ref{tab:final-values}. 

To find the uncertainties in these occurrence rate estimations we must consider the possible variations in the parameters used to calculate $\mathcal R_{\text{sBH}}$. To account for the uncertainty in the rate of critical encounters, we use Poisson statistics ($N \pm \sqrt{N}$). Following \cite{Tagawa20}, $r_{\text{AGN,MW}}$ has an allowed range of $0.06 - 0.2$ pc and $\eta_{\text{n,BH}}$ a range of $0.002 - 0.02 M_\odot^{-1}$. For the allowed range of $t_{AGN}$ we follow \cite{2018ApJ...866...66M} and use 1 - 100 Myr. With these uncertainties we get a final range of occurrence rates of 0.21 - 837 $\text{Gpc}^{-3} \text{ yr}^{-1}$ for the drag prescription $>1 M_\odot$. The results for the other prescription and GW threshold are shown in Table \ref{tab:final-values}. 

\begin{table*}[htp]
\centering
\hspace*{-1.5cm}\begin{tabular}{lccrr}
\hline\hline
%\vspace{5pt}
Migration Type & Peak GW Energy & Expected Rate  & Lower Bound & Upper Bound \\ 
               & [$M_\odot$] & [$\text{Gpc}^{-3} \text{ yr}^{-1}$] 
               & [$\text{Gpc}^{-3} \text{ yr}^{-1}$] 
               & [$\text{Gpc}^{-3} \text{ yr}^{-1}$]  \\
\hline
%\vspace{5pt}
Drag           & $>1 M_\odot$   & 3.2                  & 0.21        & 837      \\ %\hline
Drag           & $>3 M_\odot$   & 1.4                  & 0.08        & 368      \\ %\hline
Trap           & $>1 M_\odot$   & 4.7                  & 0.31        & 1194     \\ %\hline
Trap           & $>3 M_\odot$   & 1.8                  & 0.11        & 475   \\ 
\hline
\hline
\end{tabular}
\caption{Collection of the rate calculations made in this paper of close encounters expected to be detectable by LIGO. The values are organized by the migration prescription used in the population synthesis in section~\ref{sec:pop-synth} and by the peak GW energy of the encounters. The "Expected Rates" are the values obtained by using the exact proportions of critical close encounters from the population synthesis and the expected values for the AGN environment. The bounds are calculated with the uncertainties for these quantities. 
}
\label{tab:final-values}
\end{table*}

\section{Gravitational-wave Signatures}
\subsection{{Waveform Properties}}
Several waveform calculations for \gw{} from hyperbolic black hole encounters can be found in the literature. 
These calculations rely on numerical methods 
 \cite[]{Damour:2014afa,Bae:2017crk,Healy:2009zm,Levin:2008ci},
  perturbative expansions
 \cite[]{Cho:2018upo, Capozziello:2008mn, 1978ApJ...224...62K, Vines:2017hyw}, or analytic effective-one-body approaches ~\cite[]{Nagar:2018zoe,Ramos-Buades:2021adz}.
While numerical relativity simulations are accurate and computationally expensive, perturbative and effective-one-body approaches are computationally efficient but do not incorporate effects common to AGN-like scenarios, such as gas and gravitational influence of the central supermassive black hole. 

To account for the effects described in the previous sections, here we calculate waveform templates from first principles, using a finite difference method to compute the rate at which \gw{s} are emitted from the system, following the prescriptions of General Relativity. 
Although this approach is not as accurate as a full numerical relativity simulation, it provides a qualitative understanding of the time evolution of gravitational radiation in a two-body system within a realistic Jacobi framework, where two stellar mass black holes in the vicinity of a supermassive black hole \citep{Boekholt23}. 
Although this approach is a useful approximation, it introduces several sources of error compared to numerical simulations: (1) the method uses a simplified expression for the quadrupole moment, which may not fully account for higher-order multipole moments or the detailed dynamics of the binary system~\cite[]{Poisson2014}; (2) the method might not incorporate all relativistic corrections and higher-order terms that are included in numerical simulations~\cite[]{Blanchet2014}. This can lead to discrepancies, especially in strong gravitational fields, or for systems with significant relativistic motion.

According to General Relativity, the gravitational-wave strain $h$ emitted by a system of massive objects is described by Einstein's quadrupole formula: 
\begin{equation}
h = \frac{2 G}{c^4 D} \frac{d^2 Q (r, t) }{d t^2}
\end{equation}
where $D$ is the luminosity distance and $Q(r, t)$ is the quadrupole moment of the system as a function of the orbital separation between the two black holes $r$ and time $t$. 
$Q$ is approximately 
equal to the product of the binary system  total mass $M=m_1+m_2$ 
multiplied by the square of the separation \( r \) between the two black holes. This result will be accurate as long as the \gw{} wavelength is much larger than the separation $r$ between the black holes.  The timesteps have a separation of about $1.35 \times 10^{-11}$, yielding a numerical error of order $O(10^{-30})$, less than computer precision error. 

\begin{figure} [htp]
    \includegraphics[width=\columnwidth]{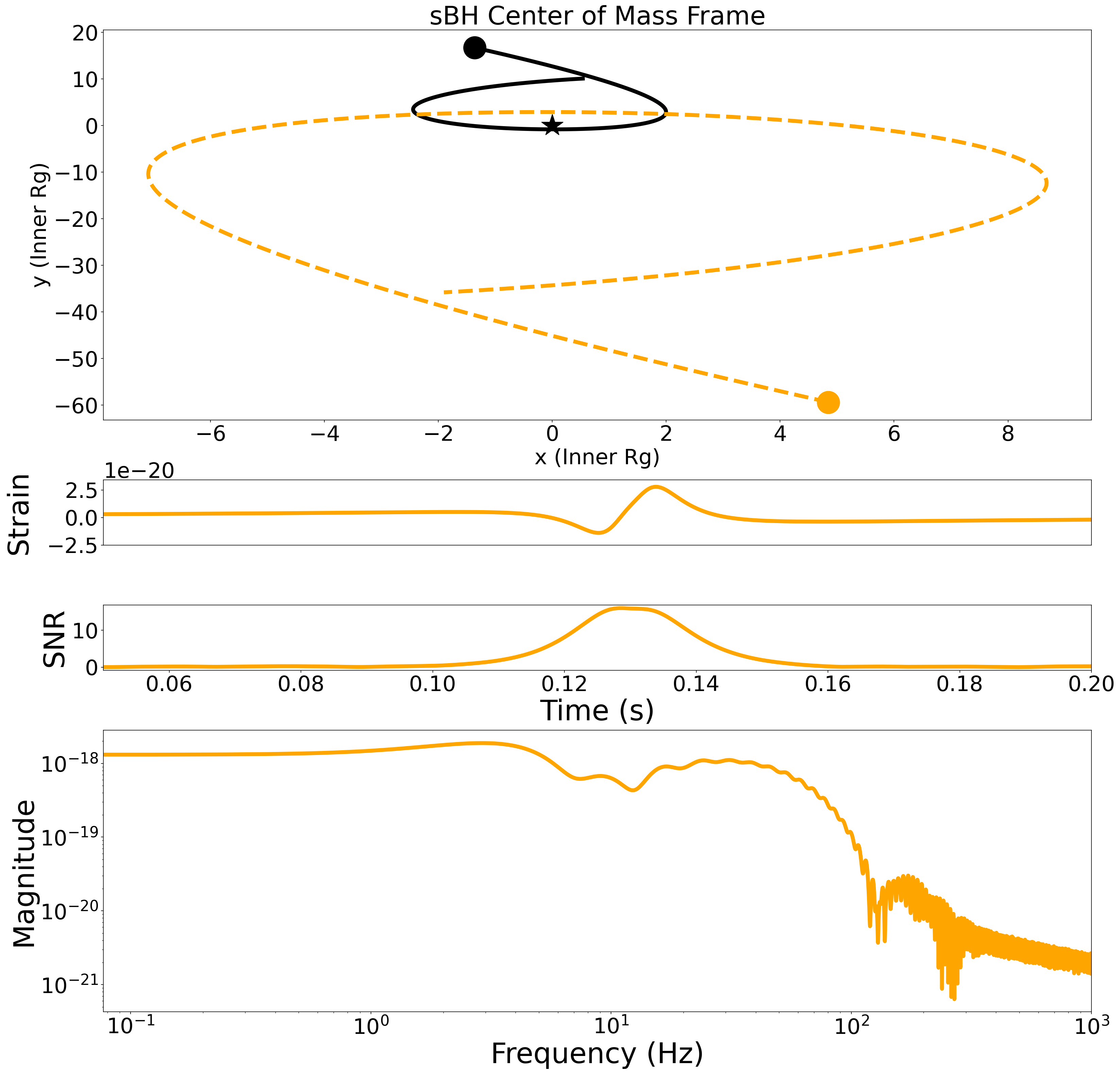}
    
     \caption{Selected \gw{} templates based on simulated trajectories for the binary with initial parameters $(\iota_1, \iota_2) = (0.000572, 0.000462)$, mean anomaly (in radians) $(\alpha_1, \alpha_2) = (4.19, 5.96)$ and initial eccentricity $(e_{init,1}, e_{init,2}) = (0.033, 0.032)$. Top panel is top-down view of black hole trajectory of primary (black) and secondary (orange, dashed) black holes in the x-y plane as computed by \rebound. The binary masses are $m_1 = 92 \;\msol$, $m_2 = 26\;\msol$, migration constant is $10^6$ The middle panel is the \gw{} strain calculated using a spline-fitting and differentiation method projected at luminosity distance $D = 10\;$Mpc. The third panel from the bottom is the matched-filter SNR as a function of time, as it would be seen in LIGO detector with zero transient noise. The fourth panel is the power spectrum as a function of frequency. The power spectrum peaks at near 5 Hz, making this one-off fly-by scenario more relevant for future ground-based detectors. } 
  \label{fig:GW_temp1}
\end{figure}

\begin{figure} [hbt]
    \includegraphics[width=\columnwidth]{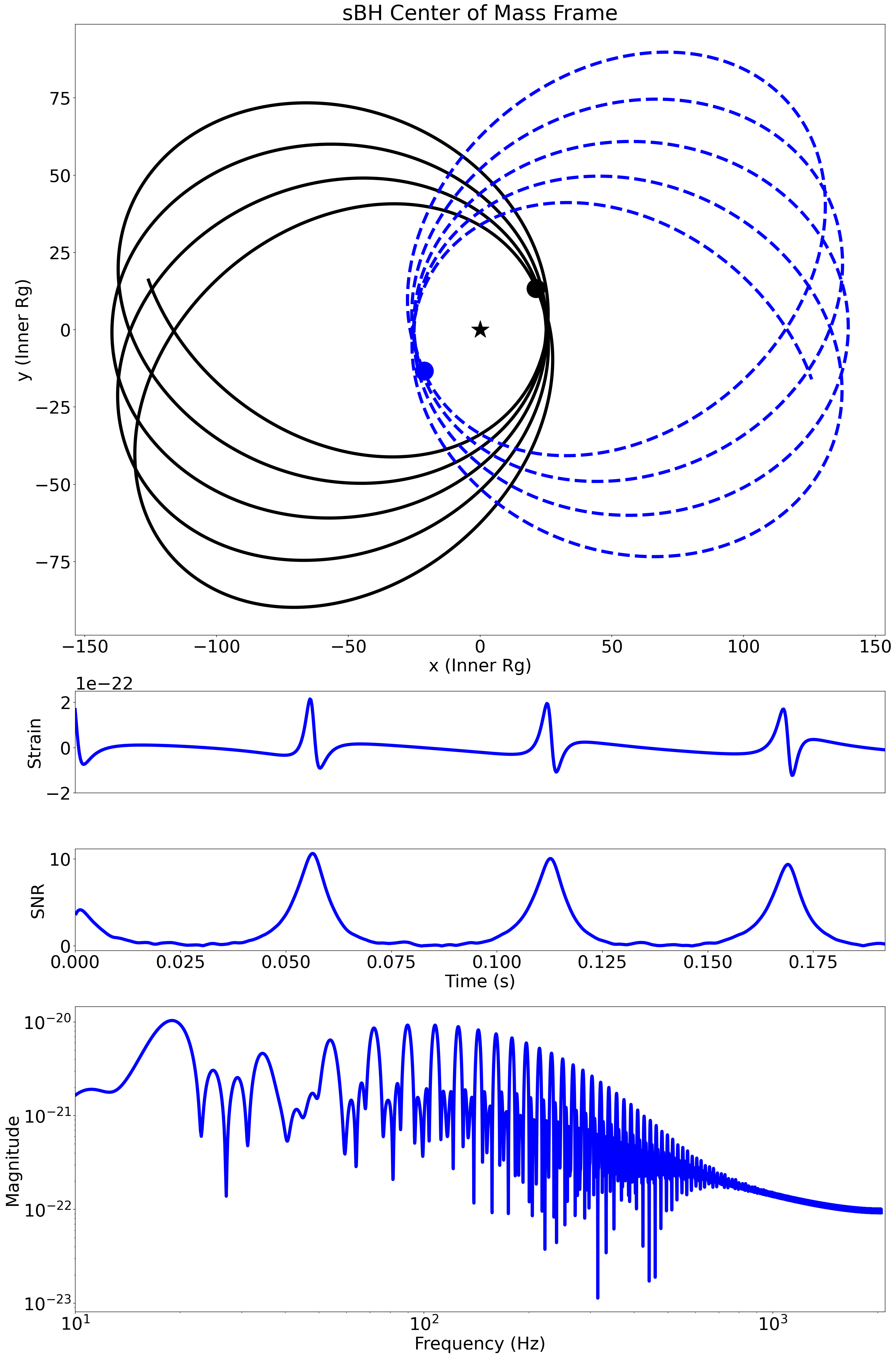}
    
     \caption{Selected \gw{} templates based on simulated \rebound{} with inclination for the binary pair $(\iota_1, \iota_2) = (0.000368, 0.000386)$, mean anomaly (in radians) $(\alpha_1, \alpha_2) = (3.60, 5.76)$ and initial eccentricity $(e_{init,1}, e_{init,2}) = (0.044, 0.023)$. Top panel is top-down view of black hole trajectory of primary (black) and secondary (blue, dashed) black holes in the x-y plane as computed by \rebound{}. The middle panel is the \gw{} strain calculated using a spline-fitting and differentiation method with luminosity distance $D = 10$ Mpc, and binary masses $m_1 = m_2 = 20\, \msol$.  The third panel is the SNR as function of time, and the fourth panel is the power spectrum as a function of frequency. 
     }
 \label{fig:GW_temp2}
\end{figure}

\begin{figure}[htp]
    \includegraphics[width=\columnwidth]{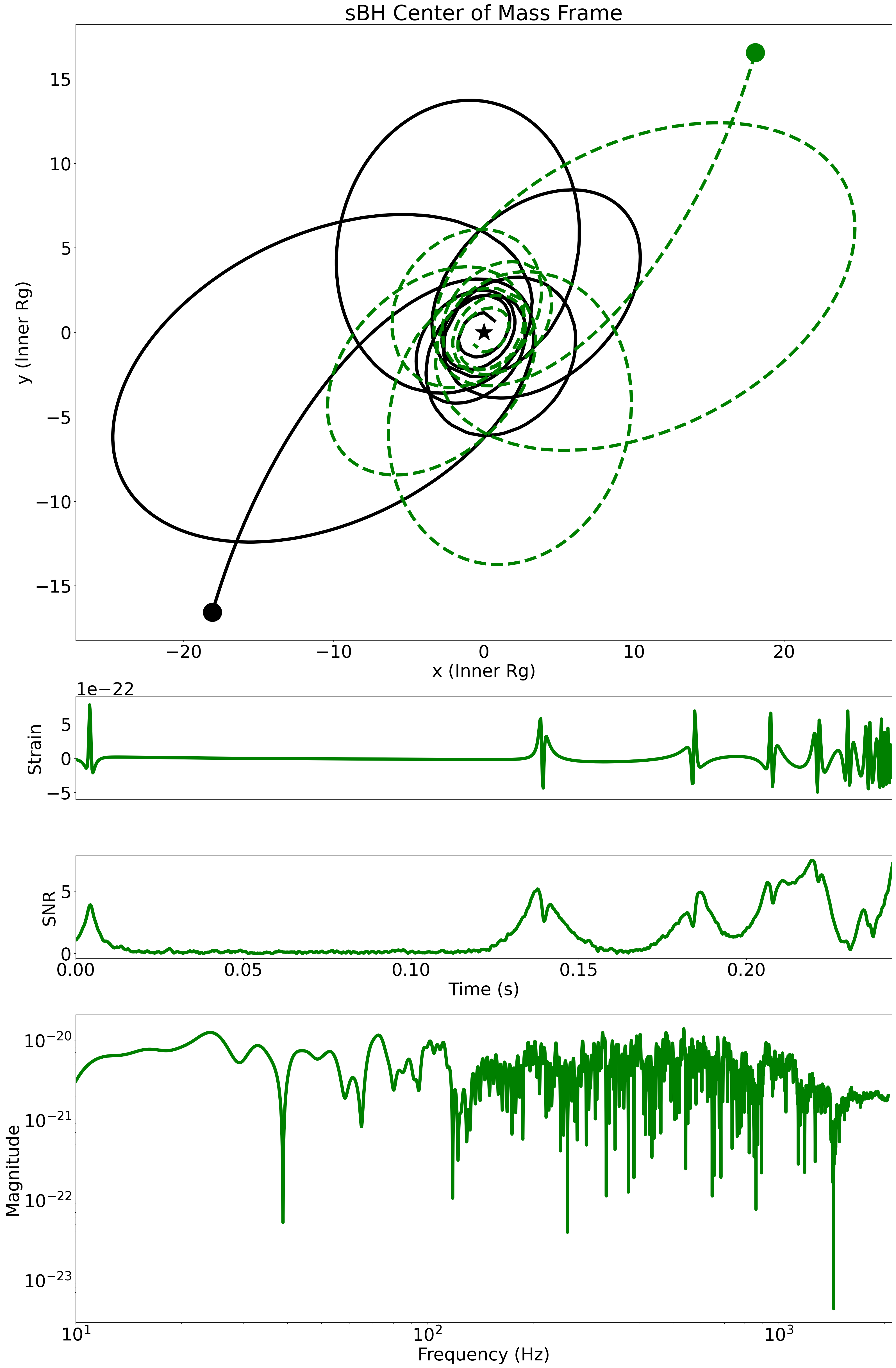}
    \centering

 \caption{Selected \gw{} templates based on simulated trajectories for the initial parameters $(\iota_1, \iota_2) = (0.000305,  0.000715)$, mean anomaly $(\alpha_1, \alpha_2) = (4.77, 4.77)$ and initial eccentricity $(e_{init,1}, e_{init,2}) = (0.039, 0.047)$. Top panel is top-down view of black hole trajectory of primary (black) and secondary (gold, dashed) black holes in the x-y plane as computed by \rebound{}. The middle panel is the \gw{} strain calculated using a spline-fitting and differentiation method with luminosity distance $D = 10$ Mpc, and $m_1 = m_2 = 20\, M_{\odot}$ solar masses. The third panel is the SNR as function of time, and the fourth panel is the power spectrum as a function of frequency. }
 \label{fig:GW_temp3}
\end{figure}

\begin{figure}[htp]
    \includegraphics[width=\columnwidth]{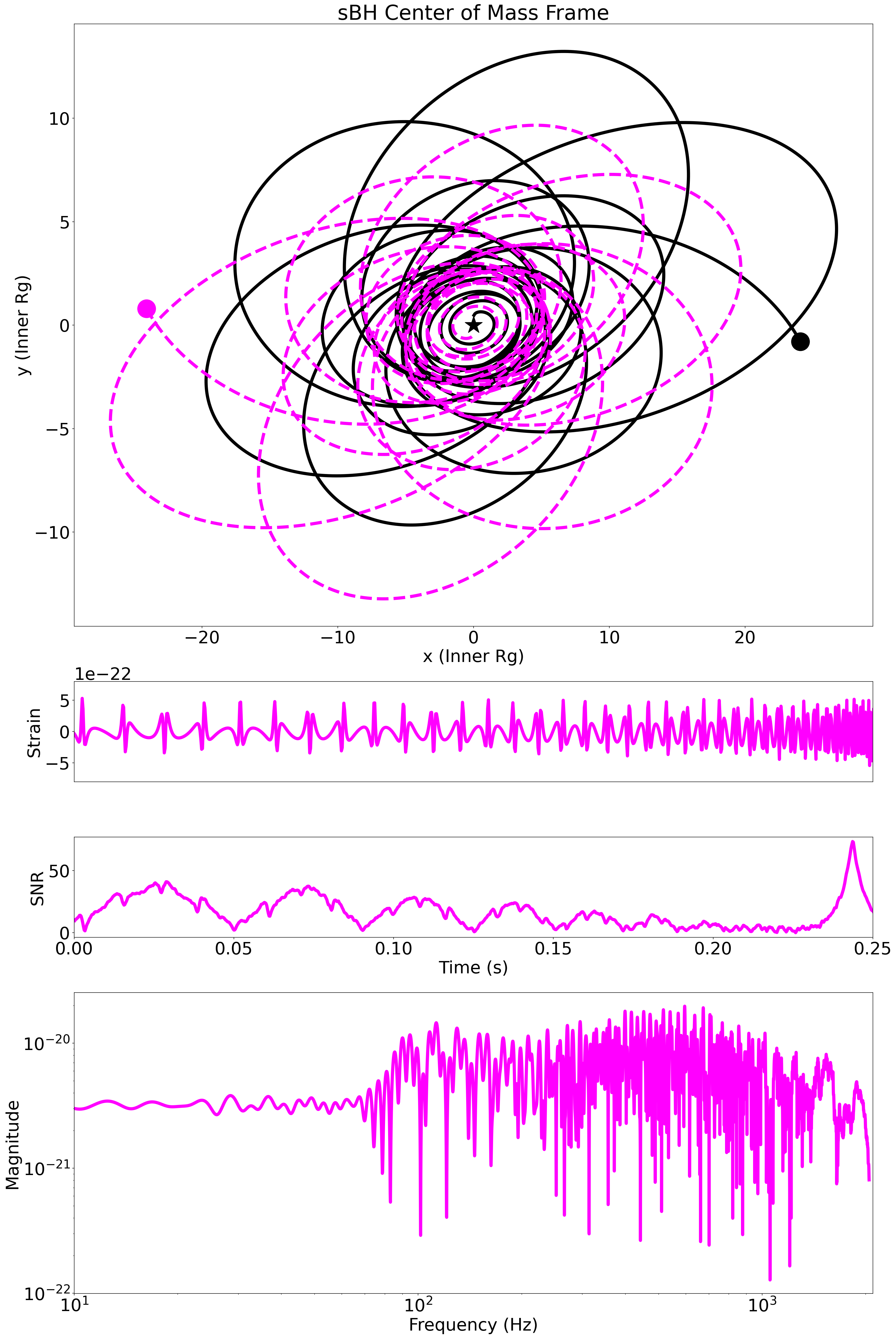}  
     \caption{Selected \gw{} templates based on simulated trajectories. $(\iota_1, \iota_2) = (0.000498, 0.000421)$, mean anomaly $(\alpha_1, \alpha_2) = (4.90, 0.89)$ and initial eccentricity $(e_{init,1}, e_{init,2}) = (0.031, 0.043)$. Top panel is top-down view of black hole trajectory of primary (black) and secondary (magenta, dashed) black holes in the x-y plane as computed by \rebound{}.  The middle panel is the \gw{} strain calculated using a spline-fitting and differentiation method with luminosity distance $D = 10$ Mpc, and binary masses $m_1 = m_2 = 20\, M_{\odot}$.  The third panel is the SNR as function of time, and the fourth panel is the power spectrum as a function of frequency. }
  
 \label{fig:GW_temp4}
\end{figure}

In Figures \ref{fig:GW_temp1} - \ref{fig:GW_temp4} we show a sample gravitational wave signatures resulting from four simulations, for illustration purposes. The initial parameters for these simulations are tabulated in Table~\ref{tab:init-values}. For each constituent black holes $m_1$ and $m_2$, Table~\ref{tab:init-values} lists the inclinations, $\iota_1$ and $\iota_2$, mean anomalies $\alpha_1$ and $\alpha_2$, and initial eccentricities $e_{init,1}$ and $e_{init_2}$, respectively. The inclination values, measured in radians, denote the tilt of the orbital planes of these bodies relative to the AGN-disk plane. Mean anomalies, also given in radians, describe the position of the bodies along their orbits relative to periapsis, measuring the time passed since the BHs last crossed periapsis. Eccentricities quantify the deviation of the bodies' orbits from circularity, with values approaching 1 indicating more hyperbolic-like orbits. For each figure referenced, specific values for these parameters are provided, allowing for comparison of the orbital characteristics as exhibited in the plotted trajectories and \gw{} signatures. The selection of these parameters is very similar to the methods used to explore the different GR prescriptions, described in section~\ref{sec:gr-prescriptions}. The inclinations are sampled from distributions designed to encourage close encounters, while the mean anomaly is sampled uniformly from 0 to $2\pi$ to best explore the possible initial conditions.

\begin{table*}[htp]
\centering
\begin{tabular}{lccrr}
\hline\hline
%\vspace{5pt}
Figure & Inclination & Mean Anomaly & Eccentricity & Impact Parameter  \\ 
       & (Inner, Perturber) & (Inner, Perturber) & (Inner, Perturber)  \\
       & [rad]       & [rad]        &  & [$R_{g,1}$]\\
\hline
%\vspace{5pt}
\ref{fig:GW_temp1}           & (0.000572, 0.000462)   & (4.19, 5.96)                  & (0.033, 0.032) & 2.65    \\ %\hline
\ref{fig:GW_temp2}           & (0.000368, 0.000386)   & (3.60, 5.76)                  & (0.044, 0.023) & 186.44   \\ %\hline
\ref{fig:GW_temp3}           & (0.000305, 0.000715)   & (4.77, 2.59)                  & (0.039, 0.047) & 0.85       \\ %\hline
\ref{fig:GW_temp4}           & (0.000498, 0.000421)   & (4.90, 0.89)                  & (0.031, 0.043) & 1.96        \\ %\hline
\hline
\hline
\end{tabular}
\caption{Collection of initial parameters of simulations and impact parameters of encounters featured in gravitational wave detection analysis. The first 3 values are for the very first time-step of each simulation. All simulations use the drag migration prescription with a time constant of $10^6$. The impact parameter is calculated at the point of closest approach of the encounters, expressed in terms of the gravitational radius of the inner black hole (Eq.~\ref{eq:grav_radius}). Note that in Figure~\ref{fig:GW_temp1} the inner and perturber black holes have masses 92 $M_\odot$ and 26 $M_\odot$ respectively, while all others have equal masses of 20 $M_\odot$.}
\label{tab:init-values}
\end{table*}

Note that the inclination of each black hole is very small, on the order of less than $10^{-4} $radians, so it does not significantly impact the final trajectories of the orbit.
The eccentricity influences how much faster the body moves at periapsis compared to other points in the orbit~\cite{vallado2007fundamentals}. Higher initial eccentricity leads to greater variation in speed and energy as the body moves between periapsis and apoapsis, and can lead to more complex orbital evolution.~\cite{murphy1995orbital}

For each simulation, figures \ref{fig:GW_temp1} - \ref{fig:GW_temp4} show the top-down trajectory of a \rebound{}  of black holes in the center of mass frame, the corresponding waveform calculated at fiducial distance of $10$ Mpc, and the signal-to-noise ratio (SNR) as a function of time. Figure \ref{fig:GW_temp1} shows the Fourier spectrum as a function of frequencies of the gravitational waveform on its bottom panel. We use the SNR definition used in other ground-based detector data analysis~\cite{LIGOScientific:2018mvr}, defined as an integral over the frequency f:

\begin{equation}\label{e:SNR}
\mathrm{SNR}^2 = 4\int_{0}^{\infty}\frac{\tilde{h}^* \tilde{h}}{S_n(f)}df.
\end{equation}

where $\tilde{h}$ is the Fourier transform as function of the frequencies of the gravitational waveform, * denotes its complex conjugate, and $S_n(f)$ is the \textit{power spectral density} (PSD), a function which characterizes the stationary Gaussian noise in the detector. Dividing by the PSD re-weights the noise spectrum toward frequencies where the detector is most sensitive~\cite{thorne_2002, abbott_2016}. In this work, we calculate the SNR using the PSD according to the projected noise spectrum projected for fifth observing run (LIGO A+ in~\cite{KAGRA:2013rdx}). 

To find the \gw{} strain emitted by the system over the lifespan of the simulation we use cubic spline fitting (\cite{d2be5552-a1ab-3d4c-89ec-5b32281f11b5}, and also section 3.3 of~\cite{nr1})and differentiation. 
Specifically, we first fit a smooth spline to the trajectory timeseries data of the simulations to interpolate between the discrete data points. This spline provides a continuous function which is used to represent the quadrupole moment. Next, we differentiate this spline function, specifically $|\vec{r}|^2$ with $\vec{r}=\vec{r_2}-\vec{r_1}$, to obtain the strain amplitude of the gravitational waves. This  allows  to accurately track and analyze the strain variations over the entire simulation span.

The strain amplitudes calculated are consistent with \gw{} \bremss{} waveforms seen in the literature \cite[]{Capozziello:2008mn, Morras:2021atg, Cho:2018upo},
with \gw{} peaks corresponding to local minima of the  black hole separation and of the system's quadrupole moment, since that is where rate of change of the quadrupole moment is greatest. 
Because the strain is proportional to the second time derivative of the mass quadrupole moment, the \gw{} strain may be understood in terms of the distance of closest approach (or periapsis distance) $r_p$, the semi-minor axis $b$ (also known as the \textit{impact parameter}), and eccentricity $e$.
Impact parameter, periapsis distance and eccentricity are related by:
\begin{equation}
b=\left\{\begin{matrix}
    a\sqrt{1 - e^2}, & e < 1 \\ \\
    -a\sqrt{e^2 - 1}, & e \geq 1
\end{matrix}\right.
\label{eqn:semimajor_ecca}
\end{equation}
and:
\begin{equation}
r_p=\left\{\begin{matrix}
    a(1 -e), & e < 1 \\ \\
    -a(e-1), & e \geq 1
\end{matrix}\right.
\label{eqn:semimajor_eccb}
\end{equation}
where $a$ is the semi-major axis. Equations \ref{eqn:semimajor_ecca} and \ref{eqn:semimajor_eccb} allow to describe $r_p$ in terms of impact parameter $b$ and eccentricity $e$ alone:
\begin{equation}
r_p=\left\{\begin{matrix}
    b\; \sqrt{\frac{1 - e}{1 + e}}, & e < 1 \\ \\
    b\; \sqrt{\frac{e-1}{e+1}}, & e \geq 1
\end{matrix}\right.
\end{equation}
%\begin{equation}
%    r_p = \frac{b (1 - e)}{\sqrt{(1 - e)(1 + e)}}
%\end{equation}
%or, 
%\begin{equation}
%    \frac{1}{r_p} = \frac{(1 - e)^{1/2} (1 + e)^{1/2}}{b}
%\end{equation},
%by 
%\begin{equation}
%    h_{peak} \propto \frac{1}{r_p} = \frac{(1 - e)^{-1/2}(1 + e)^{1/2}}{b}.
%\end{equation}
and:
\begin{equation}
    h_{peak} \propto \frac{1}{r_p} =  
    \left\{\begin{matrix}
    \frac{1}{b}\; \sqrt{\frac{1 + e}{1 - e}}, & e < 1 \\
    \frac{1}{b}\; \sqrt{\frac{e+1}{e-1}}, & e \geq 1
    \end{matrix}\right.
\end{equation}
Thus, the peak \gw{} emission  may be expressed as a relation between eccentricity $e$ and impact parameter $b$.
For scale, a luminosity distance $D=10\,$Mpc, a total mass $M = 100\,\mathrm{M}_{\odot}$, and $e = 0.99$ and $b = 2 \times 10^{-7}$ AU (equivalent to $9.72 \times 10^{−13}$ pc or $6.7 \times 10^{-2} $ $R_{g,1}$) %gravitational radii)
yields 
%\begin{equation}
   $ h_{peak} \approx  4.44 \times 10^{-21}.$
%\end{equation}
%
This suggests 
that, for constant binary mass and luminosity distance, smaller impact parameter $(O(10^{-7})$ AU), eccentricity close to unity, and  total mass in the range $40 - 120 M_\odot$ 
 increase the likelihood of  \gw{} \bremss{} within the strain sensitivity with ground-based interferometers. 

For the sake of examining only \gw{} \bremss{} one can assume $e \sim 1$ for astrophysical sources~\cite[]{Kocsis_2006}. Based on simulations in this work, $b$ falls in the range $[1.68 - 442] \times 10^{-7}$ AU (equivalent to $[8.14 - 116] \times 10^{-13}$ pc or [$0.85 - 186.44] \ R_{g,1}$) for simulations resulting in observable gravitational signatures. Note that these values for the impact parameter can be less than the collision radius of the sBHs (twice the sum of the gravitational radii of the sBHs, defined in section \ref{sec:parameter-space}); we attribute this to the chaotic dynamics of the 3-body system affecting the orbital parameters of the sBHs.

\subsection{Implications for \gw{} Searches} 
The simulations from this study broadly fall into two categories: (1) those whose binaries are caught in mean motion resonance (MMR) or avoided close-encounters as they migrate pass each other and therefore do not emit \gw{s}, and (2) those generating in \gw{} transients 
or {\em bursts}. 
The MMR depends on a functional relationship between total mass and mass ratio as discussed in section 3.2, while those that passed MMRs and avoided close encounters depend on their orbital phase and the dynamics is more chaotic in general.  

We split the \gw{} bursts further into two categories (2i) one-off flybys and (2ii) and \textit{repeated bursts}. A repeated \gw{} burst is a source which emits \gw{s} over an extended periods of time (e.g., as long as several minutes, days or even months~\cite[]{KAGRA:2022dwb}). We discuss implications for GW searches in ground based detectors. 

We examine the close encounters from our simulations, in particular the {\tt Fiducial} set discussed in section~\ref{sec:parameter-space}, to better understand how repeated burst encounters would be observed. We identify the local minima of separation between the sBHs in each close encounter trajectory and examine only those encounters with more than one local minima, i.e. repeated bursts. Of the 8,822 close encounters that occurred throughout the {\tt Fiducial} set, 44 ($\sim0.5\%$) contained repeated bursts.Thus, the rates of flyby encounters given in Section \ref{sec:pop-synth} may be adjusted accordingly to ascertain the (significantly lower) rates of repeated bursts.

\begin{figure}[ht]
    \centering
    \includegraphics[width=\columnwidth]{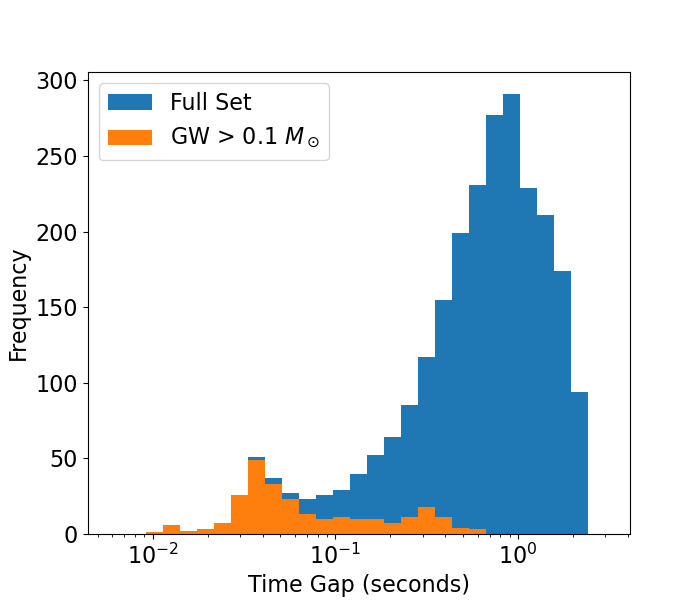}
    \caption{Distribution of time gaps between consecutive local distance minima in close encounters in the {\tt Fiducial} simulation set, with \& without restriction on the local minima for a minimum GW energy emission of $0.1 \msol$. The percentage of gaps greater than 0.1 seconds is 91.3\% in the full set and 32.6\% in the GW energy restricted set.}
    \label{fig:time-gap}
\end{figure}

Figure \ref{fig:time-gap} plots the histogram of the time gaps between the bursts. The blue color represents the full sample, and the orange color represents the sample after filtering out bursts that emit GW energy lower than $0.1 M_\odot$, as higher energy emissions are more likely to detected. It shows that time between the bursts are typically very short: most of the gap time center around $\sim 0.5$ sec when including all of the bursts, and the time gap is even shorter (centered around $\sim 0.03-0.04$ sec) when including only the bursts with high GW emission above $0.1 M_\odot$. This is because the sBH first emit low energy bursts as they migrate close to each other. They still have relatively large separation between each other, and thus have less frequent encounters  and longer time gaps between bursts. Then, as the sBHs emit more GW energy and are more closely separated between each other, they have more frequent encounters and emit more GW during each burst. Thus when keeping only those with high GW emission, the time gaps between the bursts are even shorter. We calculate the proportion of time gaps in Figure~\ref{fig:time-gap} that are greater than 0.1 seconds, as this resolution is better suited for detection. For the full set of time gaps this is 91.3\% and for GW energy $> 0.1 M_\odot$ this is 32.6\%. 

We see some correlation between the total mass of sBHs and their mass ratio with the total length of the event; in particular, larger total mass and mass ratios closer to unity correspond to longer events. This is because larger sBH masses correspond to farther separation between the sBH at a given gravitational radii, and they take longer time for encounters. Applying the GW $> 0.1 M_\odot$ restriction we see a decrease in total length between detectable minima, as this restriction will in general cut out the edges of close encounters and focus on the higher intensity periods of closer separation. 

As of the writing of this paper, there are two searches of gravitational interferometer for \gw{} \bremss{} of type (2i).
\cite{Morras:2021atg} studied 15.3 days of data from the second observing run (O2) of the LIGO-Virgo-Kagra collaboration using a neural network approach, and~\cite{Bini:2023gaj} analyzed the second half of the third observing run (O3b) with a customized version of the model-agnostic analysis pipeline coherent WaveBurst~\cite[]{Drago:2020kic}.
Both analyses use a weakly modelled search approach with minimal assumptions about the waveform, with a veto step to reject for glitches in the data. ~\cite{Morras:2021atg} found  8 candidates that are consistent with the margin of error calculated using the same algorithm. 
~\cite{Bini:2023gaj} found no significant events in O3b. Both results are consistent with the previous studies  ~\cite[]{10.1093/mnras/stab2721, Kocsis_2006, Capozziello:2008mn}, which suggest that one-off fly-by encounters are more relevant for future ground-based detectors such as Cosmic Explorer~\cite[]{Evans:2021gyd}, and Einstein Telescope~\cite[]{Maggiore:2019uih}. The sensitivities of these detectors at $10$ Hz are greater by nearly two orders of magnitude~\cite[]{Evans:2021gyd}, and therefore more sensitive to the low-frequency spectrum of the hyperbolic fly-by case.
The waveforms in figures \ref{fig:GW_temp1} - \ref{fig:GW_temp4} have a frequencies within the $0 - 2048$ Hz range employed in \gw{} searches. However, the frequency spectrum in Fig. \ref{fig:GW_temp1} peaks around the $\sim 5Hz$, revealing a frequency range more relevant for future ground-based detectors such as Cosmic Explorer and Einstein Telescope. This observation is supported by several studies ~\cite[]{Capozziello:2008mn, Kocsis_2006, Morras:2021atg, Bini:2023gaj},  suggesting that the ability to detect one-off flybys will improve in the future due to the lower frequency range. Currently, LIGO uses a frequency cutoff of 20 Hz for compact binary coalescence (CBC) events~\cite[]{LIGOScientific:2018mvr, LIGOScientific:2020ibl, KAGRA:2021vkt}. 
Figures \ref{fig:GW_temp2},  \ref{fig:GW_temp3},  and \ref{fig:GW_temp4} show frequencies around 100 Hz, making them well-suited for current ground-based detectors.
The frequencies of the waveforms align well within the operational range for continuous-wave searches~\cite[]{KAGRA:2022dwb, Wette:2023dom}, suggesting that \gw{} phenomena could be of interest for continuous-wave gravitational-wave detectors. The compatibility indicates that the waveforms may be able to be characterized using existing techniques in gravitational-wave astronomy.

Model-agnostic, wavelet based pipelines have proven to be a useful probe of \gw{} data, both as a consistency check with template pipelines~\cite[]{LIGOScientific:2021sio} and for detection of signals by their own merit, as in the case of GW15091, an IMBH detected by coherent WaveBurst with high confidence~\cite[]{Szczepanczyk:2020osv}. Unmodeled pipelines are valuable for evaluating small time segments containing astrophysical signals~\cite[]{Cornish:2014kda, Cornish:2020dwh}, providing insight into their differentiability from glitches of short-duration~\cite[]{2016PhRvD..93b2002K}, and as a consistency check with and to measure deviations from to determine deviations from GR~\cite[]{Ghonge:2020suv}. 
\gw{} \bremss{} radiation of the multiple-peak zoom whirl type, may be an interesting test case for unmodelled burst searches due to their rich time-frequency structure.
This scenario has been employed in current waveform models~\cite[]{Nagar:2018zoe, Ramos-Buades:2021adz} available in open software to interpret \gw{} waveforms in ground-based detector data.

%%%%%%%%%%%%%%%%%%%%%%%%%%%%%%%%%%%%%%
\section{Summary and Conclusion}
%%%%%%%%%%%%%%%%%%%%%%%%%%%%%%%%%%%%%%

In this paper we explored the scattering of sBHs in  AGN disks and we calculated the rate of GW \bremss{} radiation emitted in hyperbolic sBH encounters. This scenario may naturally arise in an AGN disk where migration and trap forces lead to gathering and scattering of sBHs. In such a situation, sBHs may experience high eccentricity encounters and emit GW \bremss{} radiation. 

We first examined how the parameters of the simulation affect the occurrence rate of close encounters. We determined that the implementation of first order post-Newtonian (1-PN) corrections may suppress the number of sBH close encounters, so we included both 1-PN and 2.5-PN in the simulations. We then varied the properties of the systems, concluding that near-coplanar sBHs result in more frequent and more intense encounters, and close encounters are more common around lower mass SMBHs. Moreover, we found that sBHs can be captured in first-order MMRs in the AGN disk, and this can lead to a suppression of sBH encounter rates at specific total mass of the sBH binaries. However, we note that turbulence and other orbiters in the AGN disk (not included in the simulations) may kick the sBHs out of the MMR resonances. Moreover, we note that we adopted a simple prescription for AGN disk migration, where the migration rate is not mass-dependent. Simulations with more realistic disk migration is beyond the scope of this study.

Next, we estimated event rates for \gw{} \bremss{} fly-by orbits in a larger population of systems. Using the typical parameters of AGN disks we calculate the rate of encounters that could  be detected by LIGO to be 0.21 - 837 $\text{Gpc}^{-3} \text{ yr}^{-1}$ for the drag prescription with peak GW energy release $>1 M_\odot$, and 0.31 - 1194 $\text{Gpc}^{-3} \text{ yr}^{-1}$ for the trap prescription $>1 M_\odot$ (Table~\ref{tab:final-values}). 
These rates are larger than those expected for sBH mergers: \cite{Tagawa20} adopted 1-dimensional N-body simulations and estimated a merger rate of approximately $0.02 - 60 \text{ Gpc}^{-3} \text{ yr}^{-1}$ for sBH binaries formed and evolved in the AGN disk.  \cite{Secunda2020} estimated the merger rate to be $0.66 - 120 \text{ Gpc}^{-3} \text{ yr}^{-1}$. 
This result is consistent with merging being a stricter condition than a close encounter. 

Finally, we presented \gw{} templates based on the trajectories calculated from the simulations. Using the quadrupole approximation, we estimate the waveforms from binaries emitting \gw{} \bremss{} as would manifest in ground-based \gw{} antennae. These are the zoom-whirl and fly-by orbits associated with the waveforms. Although waveforms of this type have not yet been reported in the \gw{} catalogs, we demonstrate that their \gw{} emission has an amplitude and frequency detectable by \ligo{} and other ground-based detectors, with a frequency spectrum that falls within the observable range. Furthermore, future detectors such as Cosmic Explorer and the Einstein Telescope will offer even greater sensitivity, making them well suited for detecting these types of phenomena.

\begin{acknowledgments}
The authors thank Zolt\'{a}n Haiman, Jiaru Li, Dong Lai, Amy Secunda, Saavik Ford, and Barry McKernan for inspiring discussions. GL is grateful for the support by NASA 80NSSC20K0641 and 80NSSC20K0522. PL and LC acknowledge the support of NSF grant PHY-2110481. CF is grateful for the support of NSF REU Grant 1852519. This work used the Hive cluster, which is supported by the National Science Foundation under grant number 1828187.  This research was supported in part through cyber-infrastructure research resources and services provided by the Partnership for an Advanced Computing Environment (PACE) at the Georgia Institute of Technology, Atlanta, Georgia, USA. 

\end{acknowledgments}

\bibliography{reference}{}
\bibliographystyle{aasjournal}

\end{document}